%% file: preprint2_sissa.tex
\documentclass[11pt]{report}
\usepackage{multicol}
\usepackage{apj}
\newcommand {\ks} {km~s$^{-1} \;$}

\def\lesssim{\mathrel{\hbox{\rlap{\hbox{\lower4pt\hbox{$\sim$}}}\hbox{$<$}}}}
\def\gtrsim{\mathrel{\hbox{\rlap{\hbox{\lower4pt\hbox{$\sim$}}}\hbox{$>$}}}}

\begin{document}

\bigskip
\bigskip
\bigskip
\Large
\begin{center}
{\bf THE NEARBY OPTICAL GALAXY SAMPLE: \\
THE LOCAL GALAXY LUMINOSITY FUNCTION} \\
\end{center}
\vspace{1cm}

\normalsize
\begin{center}
{\bf Christian Marinoni$^{1}$, Pierluigi Monaco$^{1,3}$, Giuliano Giuricin$^{1,2}$} \\ and {\bf Barbara Costantini$^{1,2}$}
\end{center}
\vspace{0.5cm}
\begin{center}
{\small
$^1$ Dipartimento di Astronomia, Universit\`a di Trieste,\\ via Tiepolo 11, 34131 Trieste, Italia. \\
$^2$ SISSA, via Beirut 4, 34013 - Trieste, Italia.\\
$^3$ Institute of Astronomy, Madingley Road, Cambridge, CB3 OHA, UK.\\}
\end{center}
\renewcommand{\thesection}{\arabic{section}}
\renewcommand{\thesubsection}{\thesection.\arabic{subsection}}

\newpage

\bigskip
\begin{center}
{\bf ABSTRACT}
\end{center}

In this paper we derive the galaxy luminosity function from the Nearby
Optical  Galaxy  (NOG)    sample,   which  is   a   nearly   complete,
magnitude--limited ($B\leq$14 mag), all--sky  sample of nearby optical
galaxies ($\sim$6400  galaxies  with  $cz<5500$  \ks). For  this  {\em
local}   sample, we use galaxy  distance  estimates based on different
peculiar velocity models. Therefore,  the derivation of the luminosity
function is carried  out using  the  locations  of  field and  grouped
galaxies in real distance space. 

The local field  galaxy luminosity function in  the  B system is  well
described by a Schechter  function with a  slope $\alpha\sim$ -1.1,  a
low normalization     factor  ($\Phi^{*}\sim0.006\;Mpc^{-3}$), and   a
particularly bright characteristic    magnitude   ($M_B^{*}\sim-20.6$)
($H_0= 75\;km^{-1}\;Mpc^{-1}$).  The   exact values of  the  Schechter
parameters  slightly depend on   the  adopted peculiar  velocity field
models. Peculiar motion effects are of the order of statistical errors
and cause    at most variations  of 0.8   in $\alpha$ and   0.2 mag in
$M_B^{*}$.   Our  $M_B^{*}$-value  is brighter  by   a  few tenths  of
magnitude than previous  corresponding values,  because, referring  to
total corrected blue magnitudes, better represent the galaxy light. 

Also the  selection  function, evaluated  in terms   of the luminosity
function,  appears to  be  little sensitive   to the  adopted peculiar
velocity field models,  which, however,  bear a  large impact on   the
local galaxy density on the smallest scales. 

The shape  of the  luminosity function   of spiral galaxies  does  not
differ  significantly from that of  E-S0 galaxies.  On the other hand,
the late-type   spirals  and irregulars have    a very steeply  rising
luminosity function towards the  faint end ($\alpha\sim$-2.3 -- -2.4),
whereas the  ellipticals  appreciably decrease  in  number towards low
luminosities. 

The  presence of  galaxy  systems in the   NOG sample does  not affect
significantly    the  field   galaxy     luminosity function,    since
environmental effects  on the total  luminosity  function appear to be
marginal.  The  luminosity function  of  the  members of  the  richest
galaxy systems tends to show  a slightly brighter $M_B^{*}$-value than
the norm. 

In  the light of  constraints imposed  by the  observed  galaxy number
counts, the low normalization of the luminosity function suggests that
the nearby universe ($cz\lesssim$5000 km/s) examined in this paper may
be underdense by a factor $\sim$1.5. 

\vspace*{6pt}
\noindent {\em Subject headings: }
galaxies: distances and redshifts -- 
galaxies: fundamental parameters  -- galaxies: luminosity    function,   
 mass  function -- cosmology: observations

\bigskip
\bigskip

%\begin{multicols}{2}

\clearpage

\begin{center}
{\bf 1 INTRODUCTION} 
\end{center}

Much work has gone into the determination of the local ($z=0$) optical
galaxy luminosity function (LF)  (see,  e.g., the review by  Binggeli,
Sandage  \&  Tammann 1988 for   early  works), which is  a fundamental
quantity in cosmology.  However,   there is still uncertainty  in  its
normalization and in its faint end. These quantities are important for
the question    of whether there  is  significant  evolution in galaxy
properties to  $z\sim0.5$ and for the  explanation of the faint galaxy
counts.    As  stressed in  the reviews  by  Colless  (1997) and Ellis
(1997), many  efforts are still needed  to verify the detailed form of
the local LF.  Furthermore, in  a magnitude--limited galaxy sample the
knowledge of   the LF is an essential   ingredient  for evaluating the
selection function for the sample. 
 
In this paper  we determine the  optical galaxy  LF ---and, hence  the
intimately related  selection function   --- of a   magnitude--limited
complete sample of nearby optical galaxies. The present work is indeed
meant to be an  important step towards the  recovering of  the optical
galaxy density field in the  nearby  universe (see, e.g., Marinoni  et
al.   1998a for  a preliminary account  of  this  work). As previously
discussed  (e.g., Hudson 1993, Santiago et   al. 1995), optical galaxy
samples are   more suitable for mapping   the galaxy density  field on
small scales than IRAS--selected  samples, which have  been frequently
used as tracers of the galaxy density field on large scales.

In this  work we consider the  magnitude--limited,  all--sky sample of
nearby galaxies which was  extracted by Garcia  et al. (1993) from the
Lyon--Meudon Extragalactic Database (LEDA). This sample comprises 6392
galaxies with recession  velocities $cz<$5500 \ks  and corrected total
blue magnitudes $B\leq$14 mag.   Although the different optical galaxy
catalogues, from which data are collected and homogenized in the LEDA,
have different limits of completeness in apparent magnitude or angular
diameter, the  above-mentioned  galaxy  sample    was found  to     be
substantially complete down to its  limiting magnitude $B=14$ mag, for
galactic  latitudes $|b|>20^{\circ}$)  (Garcia et  al. 1993).  On  the
basis of   the LEDA compilation of   bright galaxies with  and without
redshifts,  we estimate that   the completeness level  of this  sample
limited to  $|b|>20^{\circ}$ (5366 galaxies) is  $\sim$85\% .  For the
sample of   5832 galaxies  having  $|b|>15^{\circ}$,  the completeness
level   decreases to $\sim$80\%, according  to   the deficit of galaxy
counts at  low $|b|$-values. Most  the   bright galaxies with  unknown
redshift lie at $|b|<30^{\circ}$. 

Group assignments  for the galaxies of  this  sample have been already
provided   by Garcia (1993), who  identified   groups by employing the
percolation or {\it  friends of friends} method  proposed by Huchra \&
Geller (1982)   and the hierarchical  clustering  method  (e.g., Tully
1987). The adopted final  catalog of groups   was defined as that  one
which includes  only groups common  to the two catalogs.  Accordingly,
we shall treat 3381 ungrouped galaxies as  field galaxies. Of the 3011
members of the 485  groups  (with at  least   three members), in   the
present study we omit 22 galaxies  which are possible members (members
which almost satisfy the selection criteria). 

Garcia et  al.  (1993) tabulated  several  parameters for each galaxy,
such as the morphological type, the corrected total blue magnitude $B$
and the   corrected angular  size  (the   latters transformed in   the
standard systems of the RC3  catalog by de Vaucouleurs  et al.  1991),
and the distance modulus.  In general,  this quantity has been derived
simply  from   redshift--dependent distances  with  an  adopted Hubble
constant  $H_0=75\;km\;s^{-1}   Mpc^{-1}$,   except for    the nearest
galaxies  (with  $cz<$1500 \ks),  where redshift-independent distances
obtained  from some distance    indicators (DIs) (i.e.   from a   blue
Tully-Fisher relation which  relates luminosities to  maximum rotation
velocities and a luminosity-luminosity  index relation) have been also
partially considered. 

In   Paper    I  (Marinoni   et  al.    1998b),    we  corrected   the
redshift--distances of the  galaxies and  groups  of this sample   for
peculiar motions by means of models of the peculiar velocity field. We
purposely avoided correcting   distances by calculating  the  peculiar
velocity  field on  the basis of  the positions  and redshifts of  the
galaxy sample itself (in the linear theory approximation), because the
galaxy sample  does not presumably  include all relevant gravitational
sources (such   as   the Shapley  concentration)   for  local peculiar
motions. We  instead  followed  two  basic independent  approaches  to
modeling the  velocity  field in the  nearby  universe: i)  a modified
cluster  dipole model, which  is  Branchini \& Plionis' (1996) optical
cluster  dipole reconstruction scheme modified by  the  inclusion of a
local model  of  the Virgocentric  infall  in  the Local  Supercluster
region.  ii) a semi-linear approach which uses a multi-attractor model
fitted  to the  Mark  II and  Mark III   catalogs  of  galaxy peculiar
velocity (Willick et al. 1997).  In this  model the velocity field was
assumed to  be  generated by a   few prominent  gravitational  sources
(Virgo  cluster,    Great   Attractor,   Perseus-Pisces  and   Shapley
superclusters),  which were described  as spherically symmetric masses
with King-type mass density radial distributions. 

Furthermore, we inverted the  redshift--distance relations relative to
different velocity  field  models   and  solved the   problem  of  the
triple-valued  zones of this relation (these  zones  can appear in the
vicinity   of   prominent overdensities)   by  using blue Tully-Fisher
relations calibrated on   suitably defined samples  of objects  having
distances predicted by peculiar velocity models. 

In this way, in Paper I we provided homogeneous estimates of distances
for the   individual 3381 field galaxies  and  the 485 groups   of the
above-mentioned galaxy  sample, hereinafter denoted as  Nearby Optical
Galaxy (NOG) sample. 

There  are  three main   conceptual  differences between  our  present
analysis of the local galaxy LF and previous relevant studies based on
other,  comparably shallow samples, such  as the CfA1 (Davis \& Huchra
1982), CfA2 (Marzke et al.  1994a), SSRS2 (da Costa  et al. 1994), and
ORS (Santiago et al. 1995, 1996). 

First, we use the distance estimates given  in Paper I for determining
the galaxy LF of the NOG sample. Thus, our derivation of the galaxy LF
is carried out using  the  locations of  field and grouped  objects in
real distance space, not in  redshift space as  is usually done in the
literature.  Having calculated  different models  of galaxy distances,
we are also able to quantify to what extent differences in the current
views  on the cosmic flows can  affect the determination of the galaxy
LF of an optical galaxy sample.  We  will find that the luminosity and
selection  function  of our  optical   and  local  sample  are  little
sensitive to the   adopted  peculiar velocity field,   thus confirming
Yahil et  al.'s  (1991)  finding based   on  IRAS galaxy  samples (and
different velocity field models). 

Second,  we rely  on apparent total  B  magnitudes fully corrected for
internal extinction, Galactic  extinction and K-dimming (Garcia et al.
1993).  These corrections lead  to bright magnitudes.  In  particular,
the first correction,  which  is conspicuous in very  inclined  spiral
galaxies, is generally neglected   in generic redshift surveys,  which
comprise many faint   galaxies  of unknown inclination.   The  average
correction for generic samples of bright spirals is $\sim$0.2--0.3 mag
in the B band (de Vaucouleurs, de Vaucouleurs \& Corwin 1976; Hasegawa
\& Umemura 1993). Being selected on the basis of homogenized corrected
magnitudes, NOG is in principle designed to provide a good estimate of
the bright end of the local LF.

Third, the NOG  sample,  which  contains  a large catalog   of  groups
selected in a homogeneous   way  according to well-defined   selection
criteria, allows us  to investigate on  possible environmental effects
on the LF, specifically on  differences between the LFs of non-grouped
(field) and grouped galaxies. 

A different, complementary approach  to the definition of an  all--sky
sample of nearby optical galaxies  with good completeness in redshift,
the   "Optical  Redshift Survey"  (ORS),  was followed  by Santiago et
al. (1995).  The ORS sample contains 8286 galaxies with known redshift
and consists of two  overlapping optically--selected samples  (limited
in apparent magnitude and diameter,  respectively) which cover  almost
all the sky with $|b|>20^{\circ}$.  Each  sample is a concatenation of
three subsamples drawn from the Uppsala General Catalogue (UGC) in the
north, the European Southern Observatory (ESO) catalogue in the south,
and the  Extension to the  Southern Observatory  Catalogue (ESGC) in a
strip just south of the celestial equator.  The authors selected their
own  galaxy sample  according  to  the  raw (observed) magnitudes  and
diameters and then quantified  the effects of Galactic  extinction (as
well as random and systematic errors)  on the resulting galaxy density
field, which was calculated out  to $cz$=8000  km/s in redshift  space
(Santiago et  al. 1996). Adding the IRAS  1.2 Jy galaxy sample (Fisher
et al.  1995) in the unsurveyed  zone  of avoidance ($|b|>20^{\circ}$)
and  at   large  distances ($cz>$8000  km/s),  Baker    et  al. (1998)
calculated the resulting peculiar velocity field. 

Notwithstanding the differences in  selection criteria and  the larger
redshift  incompleteness of the NOG, the  distribution of NOG galaxies
on the sky  is  qualitatively similar  to that of  ORS galaxies;  both
samples delineate similar major structures in the nearby universe. 

The outline of   our paper is  as follows.    In \S 2   we address the
determination of the Schechter-type blue LF for the  NOG sample. In \S
3 we  describe our results concerning  the total LF, the morphological
type--specific LFs, and the  comparison between the field and  grouped
galaxy LFs.  In  \S   4  we compare  our   results with   previous  LF
determinations and discuss some implications  related to galaxy number
counts. In \S 5 we define the selection function  of the NOG sample in
terms of the galaxy LF.  Conclusions are drawn in \S 6. 

Throughout, the Hubble constant is 75 $h_{75}\;km\;s^{-1} Mpc^{-1}$. 

\begin{center}
{\bf 2 The galaxy luminosity function: analysis}
\end{center}

We evaluate the  galaxy LF of the NOG  sample in the standard B system
of  the  RC3  catalog.  We  rely  on  the  various  sets of  distances
(corrected for peculiar motions) given by  Marinoni et al. (1998b) for
field and grouped galaxies and on the total B magnitudes as derived by
Garcia et  al.  (1993), who transformed  the original raw  data to the
standard B system of the RC3 catalog.  For numerous NOG galaxies total
B magnitudes   were  carefully derived  from  aperture  photometry and
detailed (photographic  or   CCD) surface photometry.  Garcia   et al.
(1993)  corrected the total     B magnitudes for internal   extinction
(following Tully  \&  Fouqu\'e 1985),  Galactic extinction  (following
Burstein \& Heiles 1978 a,b, 1982, 1984), and K--dimming (Pence 1976). 

In general, extensive,  bright magnitude--limited samples such  as the
NOG sample, which is currently one of the  largest samples used in the
LF determination, can  provide  good determinations  of the local  LF,
except for its faint  end (i.e., roughly  for $M_B>-15$).  Compared to
other wide-angle, comparably shallow, bright magnitude--limited galaxy
samples, such   as the CfA1  (Davis  \& Huchra 1982),  CfA2 (Marzke et
al. 1994a), and SSRS2 (da Costa et al.  1994; Marzke \& da Costa 1997)
redshift  surveys,  the NOG   sample may be  less  sensitive  to local
density fluctuations because it covers a much larger solid angle. 

Deeper magnitude-limited  redshift  surveys are  in  principle  better
designed to  address issues concerning the faint  end of  the local LF
and its normalization, which   in  shallow  galaxy samples  could   be
affected  by local density   fluctuations.  But, in  practice, surveys
which extend local samples  to fainter magnitudes and higher redshifts
may begin  suffering more seriously from  some kinds of uncertainties:
i) their photometric  quality  gets  worse  and, in  particular,   the
photographic  photometry scale  may  be not  well--defined (see, e.g.,
Takamiya, Kron \&  Kron 1995, Metcalfe,  Fong \& Shanks 1995, Rousseau
et al.   1996, for  systematic errors  in  Zwicky magnitudes,  APM and
COSMOS bj photometry,  respectively); ii) their completeness gets more
doubtful,  because,  at a   given magnitude, they   tend to  miss more
compact  and low--surface  brightness  (LSB)  galaxies  by virtue   of
selection effects  inherent in  standard  image detection   algorithms
(e.g., Disney \& Philipps 1985; Davies, Disney \& Phillipps 1989).  As
a matter of fact, at  a given magnitude, galaxies  which lie at nearer
distances are selected over a  broader interval of surface  brightness
(e.g., Phillipps, Davies, \& Disney 1990). 

Much of the meaningful discrepancies between the  local LFs derived in
the literature   (see  \S 4 below)   can  be ascribed  to  photometric
problems (magnitude  scale    errors, use of  different   isophotal or
aperture magnitude systems,  lack of adequate corrections of  observed
magnitudes) and to different  observational selection criteria used in
constructing  galaxy samples.  Both  kinds  of  problems are a  little
minimized in NOG, because this sample comprises both bright and nearby
galaxies. 

Furthermore, the deepest   surveys suffer from redshift dependence  in
the  measured isophotal or   aperture magnitudes, due to the  combined
effects of the point spread function  and surface brightness variation
(e.g., Dalcanton 1998).  In   any case, very   faint magnitude-limited
redshift surveys  (e.g., Ellis 1997), which  are  very important tools
for probing  evolutionary effects on the  galaxy LF, have difficulties
in giving unambiguous constraints on the local galaxy LF, because they
are sensitive to evolutionary effects (e.g.,  the problem of the faint
blue galaxy  excess)  and to the  local normalization   problem (e.g.,
Driver \& Phillipps 1996). 

\begin{center}
{\em 2.1 Estimating the shape} 
\end{center}

We  evaluate the shape of   the  galaxy LF following Turner's   (1979)
method (see also de Lapparent, Geller \& Hucra 1989).
As recently demonstrated by Willmer (1997), who compared the robustness of
different Schechter--type LF estimators by means of Monte Carlo
simulations, this method --- as well as all widely used methods, except
the $1/V_{max}$ technique --- gives a substantially unbiased estimate of
the shape of this function, especially in the case of large samples. 

Under the  assumption that  the  LF is universal  (i.e. independent of
location in the  universe), the  galaxy number   per unit  volume  and
magnitude has a form separable into a product of functions of absolute
magnitude      $M$     and       position     $\Phi({\bf           r}
,M)dMdr^{3}=\varphi(M)\rho({\bf    r})dMdr^{3}$.

  In this  notation,
$\varphi$ represents the   fraction   of objects with   luminosity  M,
i.e. the shape of LF. 

\newpage

For a magnitude-limited sample of galaxies, the mean number of objects
with luminosity M  is

\begin{equation} \displaystyle  
n(M)dM=\Bigg[\int_{V_{max}(M)}\rho({\bf r})dV \Bigg]\;\varphi(M)dM 
=S(M)\varphi(M)dM  \label{form1} \end{equation}

\noindent where $V_{max}(M)$ is the maximum volume in which galaxies 
brighter than M are still visible.

Considering the number density of galaxies brighter than M

\begin{equation} \displaystyle  N(<M)=S(M)\int_{-\infty}^{M}\varphi(M')dM'=
S(M)\Psi(M) \label{form2}  \end{equation}

\noindent 
and taking the ratio between eq. \ref{form1}  and eq. \ref{form2},  
we are left with a quantity $Y(M)dM$ which is density-independent. 
Rewriting this quantity using the following differential form

\begin{equation} \displaystyle Y(M)dM=\frac{d\Psi}{\Psi(M)} 
\label{form3}, \end{equation}

\noindent
an analitic expression for the shape of the LF is given by
\begin{equation}\displaystyle  \varphi(M)=\phi_{0}Y(M)
exp \int_{M_0}^{M} Y(M')dM'
\label{form4} \end{equation}

\noindent 
where $M_0$ is the brightest absolute magnitude we use in the 
estimation of the shape of the LF. 

Inside $V_{max}(M)$, and for small absolute magnitude 
increments (i.e. for small distance increments $\delta r$),  
$Y(M)$ can be estimated using the following approximation 
(de Lapparent, Geller \& Huchra, 1989) 

\begin{equation} \displaystyle Y(M)\approx\frac{N(\leq M)-N(\leq M-\delta M)}
{N(\leq M)\delta M} \label{form5} \end{equation}.

\noindent where $N(\leq M)$ is the number of galaxies brighter than M. 
Even if this estimation technique is intrinsically non parametric, 
in order to make comparisons 
with previous works, and also in order to avoid biases in the determination
of the LF shape (de Lapparent et al., 1989), 
we fit eq. \ref{form5} 
against the one predicted,
assuming, for the LF shape $\varphi$ in the N($<$M) estimator 
(eq. \ref{form2}),  the Schechter (1976) form

\begin{equation} \displaystyle \varphi \propto 
 10^{0.4(M^{*}-M)(\alpha+1)}\textrm{exp}
[-10^{0.4(M^{*}-M)}]. \label{form6} \end{equation}

\noindent  
In this  equation, $M^{*}$  is a fiducial  magnitude which characterizes
the point (the "knee" of the function)  where the form of the function
changes from a power-law (with slope $\alpha$) to an exponential slope
 
With this least-square  algorithm, weighted assuming Poisson fluctuations,
we can measure the best parameters $\alpha$ and $M^{*}$ for the Schechter 
function.

However,  in order to account for  random errors in the magnitudes, we
convolve the  Schechter    function  $\Phi_{s}(M)$  with   a  Gaussian
magnitude error distribution with zero mean and dispersion $\sigma_{m}
= 0.2$ mag to give an observed LF $\Phi_o(M)$:

 \begin{equation}          \Phi_o(M)=\frac{\displaystyle1}{\displaystyle
\sqrt{2\pi\sigma_{m}}}   \int_{-\infty}^{+\infty}\;  \Phi_{s}(M')\;exp
\bigg[\frac{-(M'-M)^{2}} {2\sigma_{m}^{2}}\bigg]dM'. \label{form7} 
\end{equation}

We perform the  least-square fit of the  LF estimator Y  using various
distance intervals of $\delta   r=$250, 500 and  700 km/s.   The lower
limit for $\delta r$ was fixed by the spread caused by random noise in
the peculiar velocity field, while the upper limit was set at the peak
value of the absolute strength of the peculiar velocities. A variation
in  $\delta r$ over  the  range 250-700 \ks   causes changes in the LF
parameters  $\alpha$   and   $M^{*}$  which are    of  the    order of
uncertainties. 
\begin{center}
{\em 2.2 Estimating the normalization}
\end{center}
Involving ratios between the differential and the integrated LFs,
the estimator Y does  not provide   any  information  about  the
normalization factor $\phi^{*}$ (i.e., the scaled number density of galaxies with luminosity $M^{*}$) which has to be derived independently. We  
normalize the  Schechter function using   the fact  that the  mean
spatial galaxy density $n$ is related to the LF by the relation

\begin{equation} \displaystyle
\phi^{*}=\frac{\displaystyle  n}{\displaystyle
\int_{-\infty}^{M{s}}\;\Phi(M)dM}, \label{gio} \end{equation}

\noindent where $\Phi_o(M)$ is the convolved Schechter function 
with $\phi^{*}$ set equal to one. Since
the LF is poorly constrained at the faint end, we cut off the integral at
the limit of $L_s=L_{min}(r_s)$, where $L_{min}(r)$ is the minimum
luminosity necessary for a galaxy at distance $r$ (in Mpc) to make it into
the sample and $r_s$ is taken equal to $r_s= 500/(75\cdot h_{75})\sim 6.7
h_{75}^{-1}$ Mpc.  $L_{min}(r)$ corresponds to the absolute magnitude $M_B
= -5 \log r -25 + B_{lim},$ where $B_{lim}=14$ mag is the limiting
apparent magnitude of our galaxy sample; thus $L_s$ corresponds to $M_s=
-15.12 + 5 \log h_{75}$. 

Various methods  have been  proposed   to estimate the  mean  density.
Given our   complete   and magnitude--limited sample   of  galaxy with
$r_{min}<r_{i}<r_{max}$, we calculate the mean  galaxy density $n$
using the unbiased estimator
\begin{equation} 
fn=\frac{\displaystyle   \sum_{i}^{Ngal}\;  w(r_i)}{\displaystyle
\int_{r_{min}}^{r_{max}}\;\left( \frac{dV}{dr}   \right)dr   S(r)w(r)}
\label{dnorm},
\end{equation} 

\noindent 
where $S(r)$ is   the selection function given in   equation
\ref{selfun}  below, f=0.8 is the adopted com--
\newpage 
\noindent pleteness level (see
\S 1) and $w(r)$ is the weight function

\begin{equation} w(r)=\frac{\displaystyle 1}{\displaystyle [1+4\pi n
J_{3}(r_c)F(r)]},             \qquad             \qquad         \qquad
J_{3}(r_c)=\int_{0}^{r_{c}}\;r^{2}\xi(r)dr \label{w} \end{equation}

\noindent  which minimizes  the variance  in  the estimate of $n$
(Davis  \&  Huchra,   1982). For the  second   moment  of the  spatial
two--point galaxy correlation function  $\xi(r)$, we have adopted  the
value $J_{3}=320 h^{-3}Mpc^{3}$  obtained  for
$r_c=40\;h^{-1}Mpc$ (Marzke et al. 1994a).  Even   if the  value  of  $J_{3}$  is uncertain,because the contribution for  $r>r_c$ could be substantial,  
the final
result for $ n$ (and $\phi^{*}$) depends  little on the  value of
$J_{3}$ used (Loveday  et al. 1992). In  our case, doubling or halving
the  adopted value of $J_3$ makes  a few per  cent difference to $
n$.
  
In  this  minimum-variance  weighting scheme,  $n$  is determined
iteratively using eqs. \ref{dnorm}  and  \ref{w}   while, for errors
estimate, we add  in quadrature  the  uncertainties in $n$  arising
from  galaxy  clustering  ($\delta  n  \sim   n  [4\pi
J_{3}/V]^{0.5}$) with those arising from varying the parameters of the
luminosity function along their joint 1 $\sigma$ error ellipse (Lin et
al. 1996).

We have checked that  the incompleteness of  NOG does not  appreciably
depend  on  galaxy   morphological type and    mean surface brightness
(inside  the isophote 25 B-mag  arcsec$^{-2}$), whereas it gets weaker
at brighter B  magnitudes (e.g., $f$=0.90 (0.92) for  $|b|>15^{\circ}$
($|b|>20^{\circ}$)    and   $B\leq$13.5;      $f$=0.93   (0.96)    for
$|b|>15^{\circ}$   ($|b|>20^{\circ}$)  and $B\leq$13.0). After  having
derived in detail the degree of  completeness of NOG  as a function of
the  $B$ magnitude,    under   the simple  assumption   of  a  uniform
distribution of   the  missing galaxies  in redshift    space, we have
calculated that this  dependence  does not  appreciably affect  the LF
shape; it translates into a bias towards $M^{*}$-values systematically
brighter by 0.1 mag (and to  $\alpha$-values systematically flatter by
0.01 at most). 

In  order to  take  into account  border effects which  arise from the
mapping between redshift    space and true-distance space  and   which
influence the completeness of the sample (see Fig.  2 below),
we  evaluate the Schechter-type  LF  for the  galaxies  with distances
$r<5250/(75\;h_{75})$ Mpc only.   Further  we consider  only  galaxies
having $|b|>15^{\circ}$, in order to avoid a large incompleteness, and
$M_B$-values in the  range $-22.5 \leq M_B  - 5 \log h_{75}  \leq M_s$
(where  $M_s=-15.12$), in order  to  avoid small  number statistics in
each bin. 

We use  the  sets  of  distances $r$  relative to  different  peculiar
velocity field models,  i.e., the multi--attractor  model based on the
whole Mark  III data and  on Mark  III  spiral data (Mark  III$^{*}$),
Branchini \& Plionis' (1996)  cluster dipole model, the cluster dipole
model as modified by Marinoni et al.   (1998b) with the inclusion of a
local model of the  Virgocentric infall, the unperturbed Hubble  flows
in the  CMB and  Local   Group (LG)  frames.  The number  of  galaxies
($N\sim$ 5350)  which fulfills  the above-mentioned selection criteria
changes according to the distance model adopted. 

\clearpage

\begin{center}
{\bf 3 The galaxy luminosity function: results}

{\em 3.1 The total luminosity function}
\end{center}

Table 1 reports the Schechter  parameters obtained for sets
of  distance models.  The  errors given for  $\alpha$ and $ M_{B}^{*}$
are the  projections onto  their axes  of  the $1   \sigma$ confidence
ellipse level, as     derived from the  $\chi^{2}$   matrices   of the
least--square  fit  of Y($M_B$).

The $1 \sigma$  formal  errors  in
$\phi^{*}$ include contributions both  from galaxy clustering and from
uncertainties in $M_B^{*}$   and $\alpha$.    We have  checked    that
neglecting  magnitude errors makes  the slope $\alpha$ steeper by 0.03
and  $M^{*}$ brighter by 0.1  mag,  on average.  

The  last columns  of
Table    1    contain  the   galaxy    luminosity   density
$\rho_{L}=\int_{L_s}^{\infty}\;L \Phi(L)dL$   and  the  galaxy  number
density $n$  together with the  $1 \sigma$  errors. Letting  the lower
integration    limit   go to   zero    does   not increase  $\rho_{L}$
significantly.

Fig. 1  show  our results for the  normalized   LFs and the  $1
\sigma$ error ellipses in   the  ($\alpha$, $M_B$) plane, for   galaxy
distances predicted by the   Mark  III multi-attractor model  and  the
cluster dipole model. Fig. 1  and the values of $\chi^{2}/dof$
given in  Table   1 indicate  that the   Schechter function
always  gives a  good fit  to  the LF  of our   sample.  In the lowest
luminosity bin there might be a hint for an excess of objects relative
to the plotted fits  for the Mark III  multi-attractor models, but not
for the cluster dipole model. 

Fig. 2   displays   the observed  galaxy  number--distance
histograms   together  with the  curves   which   show  the   expected
distribution for an  uniform universe, for different  distance models.

%%TAB1
\vspace{15mm} 
\hspace{-15mm}
\begin{minipage}{18cm}
\renewcommand{\arraystretch}{1.8}
\renewcommand{\tabcolsep}{0.5mm}
\begin{center}  
\vspace{-3mm}
TABLE 1\\
\vspace{2mm}
{\sc The parameters of the Schechter LFs for different peculiar 
velocity field models.\\} 
\scriptsize
\vspace{2mm}
\input{tab1}

\end{center}
\vspace{3mm}
\label{lfdata}
\end{minipage}  

\clearpage

%%FIGURE 1%%%
%\end{multicols}
%\begin{figure}
\includegraphics{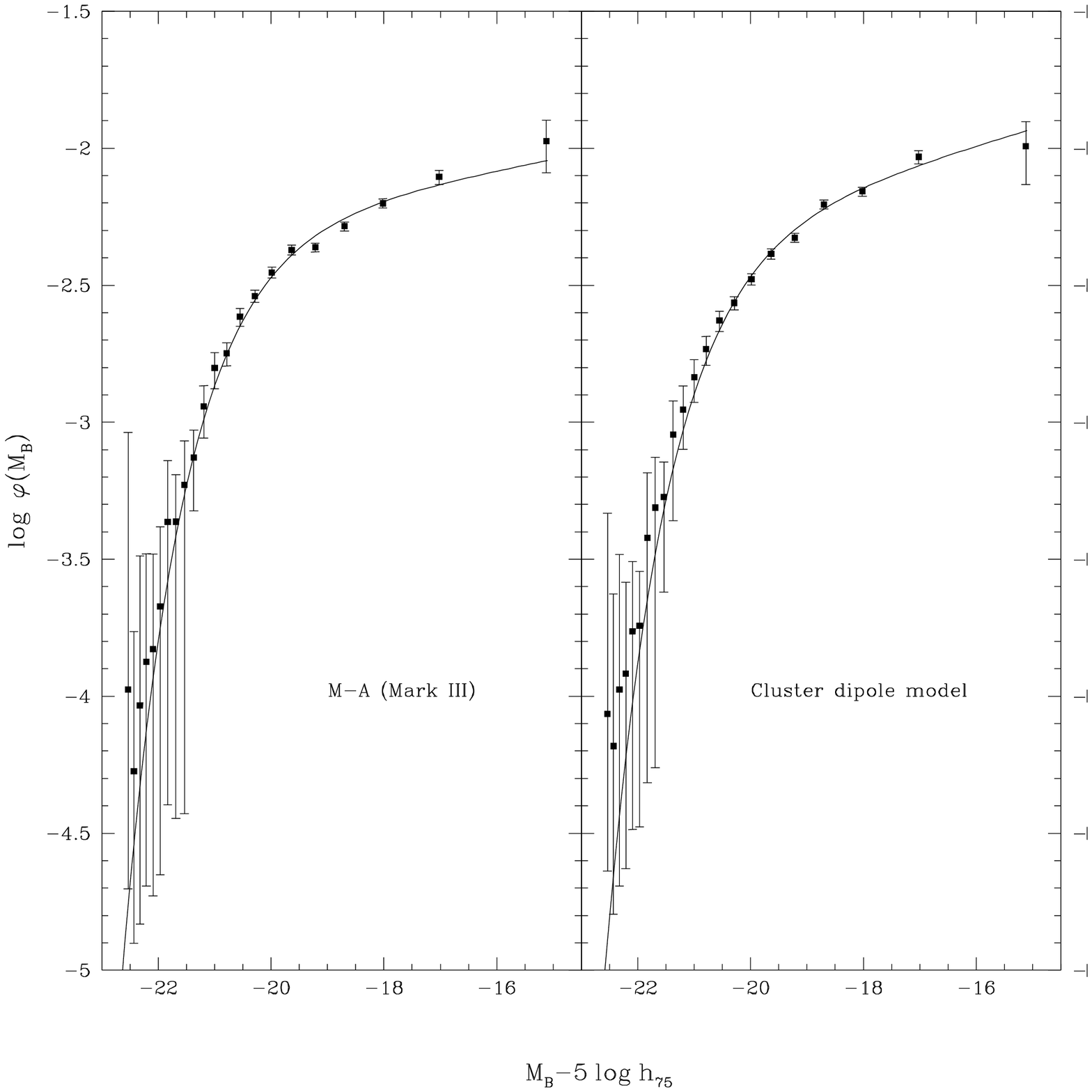}
$\ \ \ \ \ \ $\\
\vspace{18.0truecm}
$\ \ \ $\\
%\vspace{-30mm}
{\small\parindent=3.5mm {Fig.}~1.---
We     plot    the normalized differential   LFs
(together with 1$\sigma$ error bars) for galaxy distances derived from
the multi-attractor model based on all  Mark III data ({\it left}) and
from the cluster dipole model data ({\it right}).}
\vspace{5mm}
%\begin{multicols}{2}

\clearpage

%%FIGURE 2%%%
%\end{multicols}
%\begin{figure}
\includegraphics{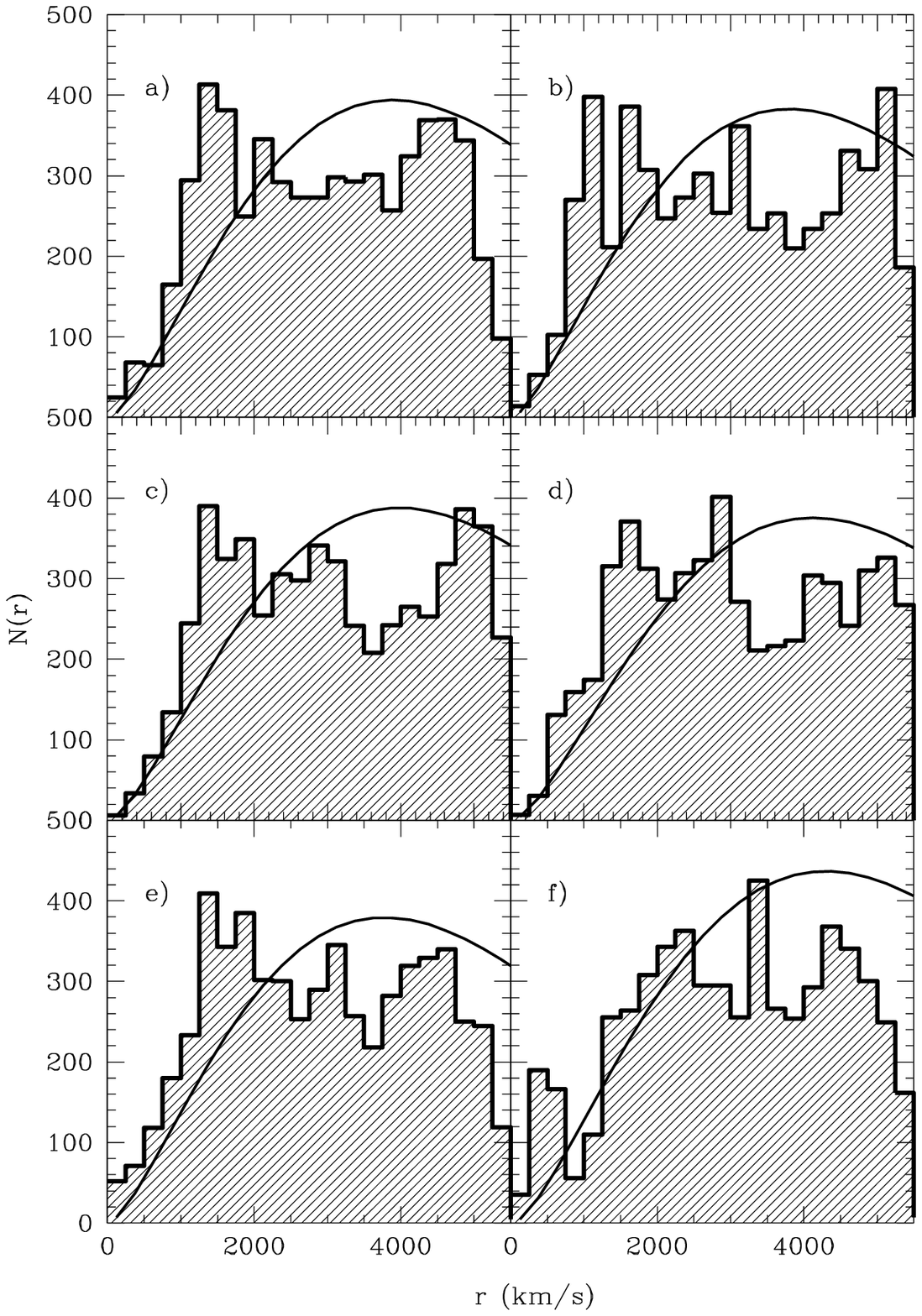}
$\ \ \ \ \ \ $\\
\vspace{17.7truecm}
$\ \ \ $\\
%\vspace{-20mm}
{\small\parindent=3.5mm {Fig.}~2.--- The  plots    show the   
observed         galaxy
number-distance  histograms. They refer to redshift-distances obtained
using  the Hubble relation in the  CMB frame (a)  and in  the LG frame
(b), and  to distances  corrected  according to  the  multi--attractor
model fitted to   the Mark III  data (c),  the multi--attractor  model
fitted to the Mark  III$^{*}$ data (d),  the cluster dipole model (e),
and the  modified cluster dipole   model (f).  The superimposed  solid
curves indicate  the predictions given  by the best Schechter function
fits in the case of a homogeneous galaxy distribution.} 
\vspace{5mm}
%\begin{multicols}{2}

\clearpage

The details of  the large "peak   and trough" fluctuations, which  are
seen  both in  wide-angle  (e.g.,  the  CfA2  sample) and narrow-angle
surveys  (e.g.,  the ESP  sample  by  Zucca  et  al.  1997)  are quite
different in the various plots. Thus, the  effects of peculiar motions
have a large  impact on  the evaluation  of the  local galaxy density,
especially on the smallest scales. On  large scales, an overdensity at
a distance of $\sim$1500--2000 \ks (corresponding to the center of the
Local Supercluster) and  an   underdensity around $\sim$3500 \ks   are
common to all histograms.

All  cases lead to  similar  values of $\phi^{*}$,  $\rho_L$, $n$, and
$\alpha$. All cases give also similar $M_B^{*}$-values, except for the
modified cluster dipole model. Owing to its large Virgocentric infall,
the modified cluster dipole model moves nearby galaxies quite close to
us (see  Fig. 2f).  Moreover,  it gives  a $M_B^{*}$--value
typically brighter by $\sim$0.1 -- 0.2 mag than the norm, in agreement
with the discussion  by Efstathiou,  Ellis  \& Peterson (1988) on  the
effect of a comparable Virgocentric infall on the LF of the RSA galaxy
sample (Sandage \& Tammann 1981). However,  in recent years there is a
growing evidence  that the Virgo cluster is  not a major source of the
peculiar velocity field in the  Local Supercluster.  From the analysis
of the magnitudes of bright cluster galaxies in a way which is free of
assumptions  about  motions in   the   LS,  Gudehus (1995)   found  no
significant  evidence of a Virgo infall  of the LG. Therefore, in what
follows we do not longer use the modified  cluster dipole model, which
is likely to be the least realistic model. 

To  sum up, our calculations   prove that differences in the  peculiar
velocity field models, though appreciably affecting the galaxy density
field  on galaxy-galaxy distance scale, do  not have  large effects on
the optical LF of a  shallow galaxy sample which  covers a very  
large solid angle.   The effects could be greater  in samples  restricted to
fairly  narrow solid angles (see, e.g.,  Davis \& Huchra 1982, for the
effect of  a  Virgocentric infall  on the  northern part  of  the CfA1
sample). 

The values  of $\rho_L$ and $n$  tabulated in Table 1 refer
to   the  population of    cataloged   galaxies  of  highish   surface
brightness. In general,  magnitude-limited  redshift surveys  miss  an
increasing fraction of LSB objects  at fainter magnitudes, which makes
it difficult  to determine the  faint end  of the  LF.  Among specific
investigations designed to recover the LSB field population (e.g., the
review by Impey  \&  Bothun 1997), the large  APM  LSB redshift survey
allowed the reconstruction of the LF of LSBs (Sprayberry et al. 1997),
which  shows that the  LSB  population (dominated by late-type spirals
and irregulars), becomes appreciably  numerous for  magnitudes fainter
than $M_B\sim$-15 only. 

Hence,   missing LSB objects do  not  affect
appreciably  the  bright   end  of  our LF.   In   the  faint end they
contribute strongly to the number density of galaxies (presumably also
to  cosmological mass density in   baryons; e.g., Bristow \& Phillipps
1994), but not too much to the  luminosity density.  Sprayberry et al.
(1997) evaluated a contribution of $0.2\;h_{75}\;L_{\odot}\; Mpc^{-3}$
(in  the B  system) from the  LF   of the  APM  LSB  galaxies, whereas
Dalcanton  et  al.   (1997a) derived  $\rho_L=0.1\;h_{75}\;L_{\odot}\;
Mpc^{-3}$ (in the  B system) from their small  sample of late-type LSB
galaxies identified in a large--area transit scan CCD data.

With  the  addition of  the  $\sim$10\% light   contribution  given by
Sprayberry et   al.   (1997), our    mean value  of  $\rho_L=2.0\pm0.3
L_{\odot}\;  h_{75}\; Mpc^{-3}$  implies  a  critical  mass--to--light
ratio $(\frac{M}{L})_{c}= \frac{1.563\cdot 10^{11}\;h_{75}}{\rho_{L}}=
7.8\pm1.2\cdot 10^{2}\;h_{75}\;(M/L)_{\odot}$   to close the universe.
The cosmological   density    parameter $\Omega_0=(M/L)/(M/L)_{c}$  is
$\sim$0.2--0.3 if we  set  $M/L\sim200\;h_{75}\cdot (M/L)_{\odot}$,  a
typical virial mass--to--light ratio estimated from galaxy clusters in
the blue (e.g., Bahcall, Lubin \& Dorman 1995;  David, Jones \& Forman
1995; Girardi et  al. 1999).  However,  on larger scales, our analysis
of cosmic flows (see  Paper I) leads to a  larger value of $\Omega_0$,
$\Omega_0\sim$0.5  (see also the  reviews by  Dekel, Burstein \& White
1997 and Hamilton 1998). 

\bigskip

\begin{center}
{\em 3.2 The type--specific luminosity functions} 
\end{center}

\bigskip
Lumping  together giants and dwarfs, we  divide  the NOG galaxies into
five   type bins, the ellipticals   (E),  lenticulars (S0), the Sa--Sb
spirals  (which cover  the range   between  S0/a and Sb),  the  Sc--Sd
spirals (which  cover the  range between  Sbc  and Sd), and the  Sm-Im
objects. We also  consider the division into two  type bins,  the E-S0
objects and the spirals and irregulars (S-Im). The morphological types
are available for almost all the NOG galaxies.

In Table 2 we give the results  for the type-specific LFs
(over the same ranges of  distances and  absolute magnitudes used  for
the  whole   sample),  for different sets    of   distance models.  In
Fig. 3 we plot the type-specific  LFs and the  total LF, in
the  specific case of Mark III  distances. In Fig. 4 We plot
the  $1\sigma$  confidence ellipses   for the parameters  $\alpha$ and
$M_B^{*}$ relative to the LFs shown in  Fig. 3 for different
morphological types.

The  morphological type--dependence of the  LF  is little sensitive to
the  distance   models  adopted.  Assembling together    the  E and S0
galaxies, we obtain a E--S0 LF which remains almost flat, although the
LF of the giant objects alone can be described  by a Gaussian 
function (e.g.,  Binggeli et al.  1998; Biviano et  al. 1995).  More precisely,
the  LF of E galaxies  appreciably declines towards faint luminosities
(with $\alpha\sim$-0.5),  whilst the LF  of  S0 objects does  not. 

The number of Sa--Sb  types declines towards  fainter magnitudes according
to  a slope ($\alpha\sim$ -0.6 --  -0.8) which is considerably flatter
than the  norm  for spirals, whilst  the Sm-Im  objects,  which are on
average,  fainter than the norm,   always exhibit  a very steep  slope
($\alpha\sim$-2.3   --  -2.4).  The  slope     $\alpha$ tends to   get
progressively steeper as  one goes from early  spirals to late spirals
and irregulars. The value of $M_B^{*}$ is fainter for S0 galaxies than
for the other types. 

\clearpage

%%%%%%%%%%%%%%%%%
%%%%%%%%%%%%%%%%%%%%%%%%%%%%%%%
%%   macro per inserire tabelle
%%TAB2
\vspace{10mm} 
\hspace{-15mm}
\begin{minipage}{17.8cm}
\renewcommand{\arraystretch}{1.4}
\renewcommand{\tabcolsep}{1.2mm}
\begin{center}  
\vspace{-3mm}
TABLE 2\\
\vspace{2mm}
{\sc The parameters of the morphological type-specific LFs.\\}
\footnotesize
\vspace{2mm}
\input{tab2}

\end{center}
\vspace{3mm}
\label{lfdatatm}
\end{minipage}  

\bigskip
\bigskip

\noindent Except for the  earliest (E) and latest (Sm--Im)  types, there  
are no
large  differences  between the shapes   of the type-specific LFs.  In
particular,  the shape of the   E--S0 LF does not differ  significantly
from that of the S--Im objects. 

\clearpage

%%FIGURE 3%%%
%\end{multicols}
%\begin{figure}
\includegraphics{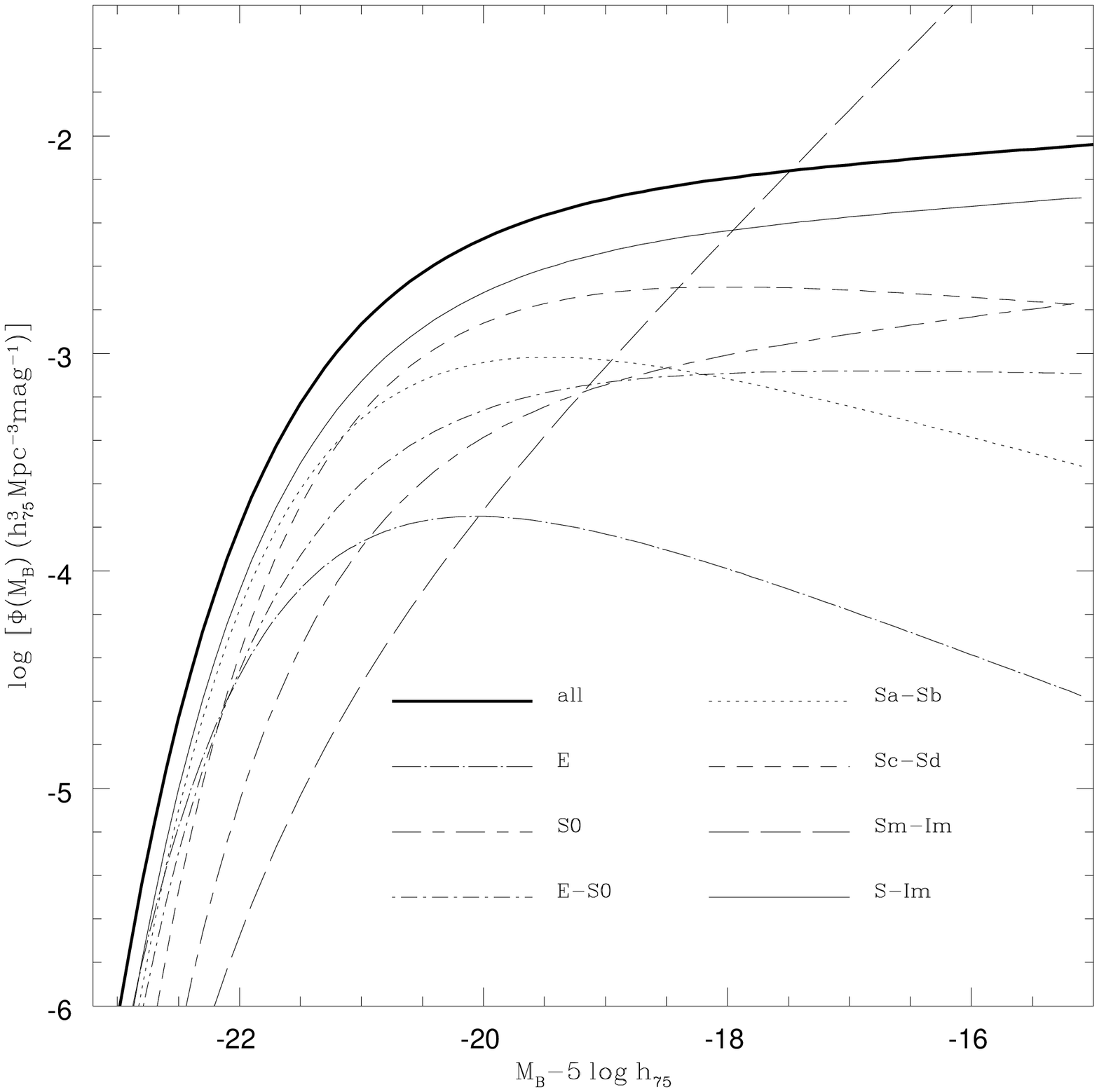}
$\ \ \ \ \ \ $\\
\vspace{18.0truecm}
$\ \ \ $\\
%\vspace{-20mm}
{\small\parindent=3.5mm {Fig.}~3.---
We plot the normalized  LFs for all galaxy types
and for different  morphological types,  adopting galaxy distances  as
given  by  the   multi--attractor  model    fitted to the   Mark   III
data.}
\vspace{5mm}
%\begin{multicols}{2}

\clearpage

%%FIGURE 4%%%
%\end{multicols}
%\begin{figure}
\includegraphics{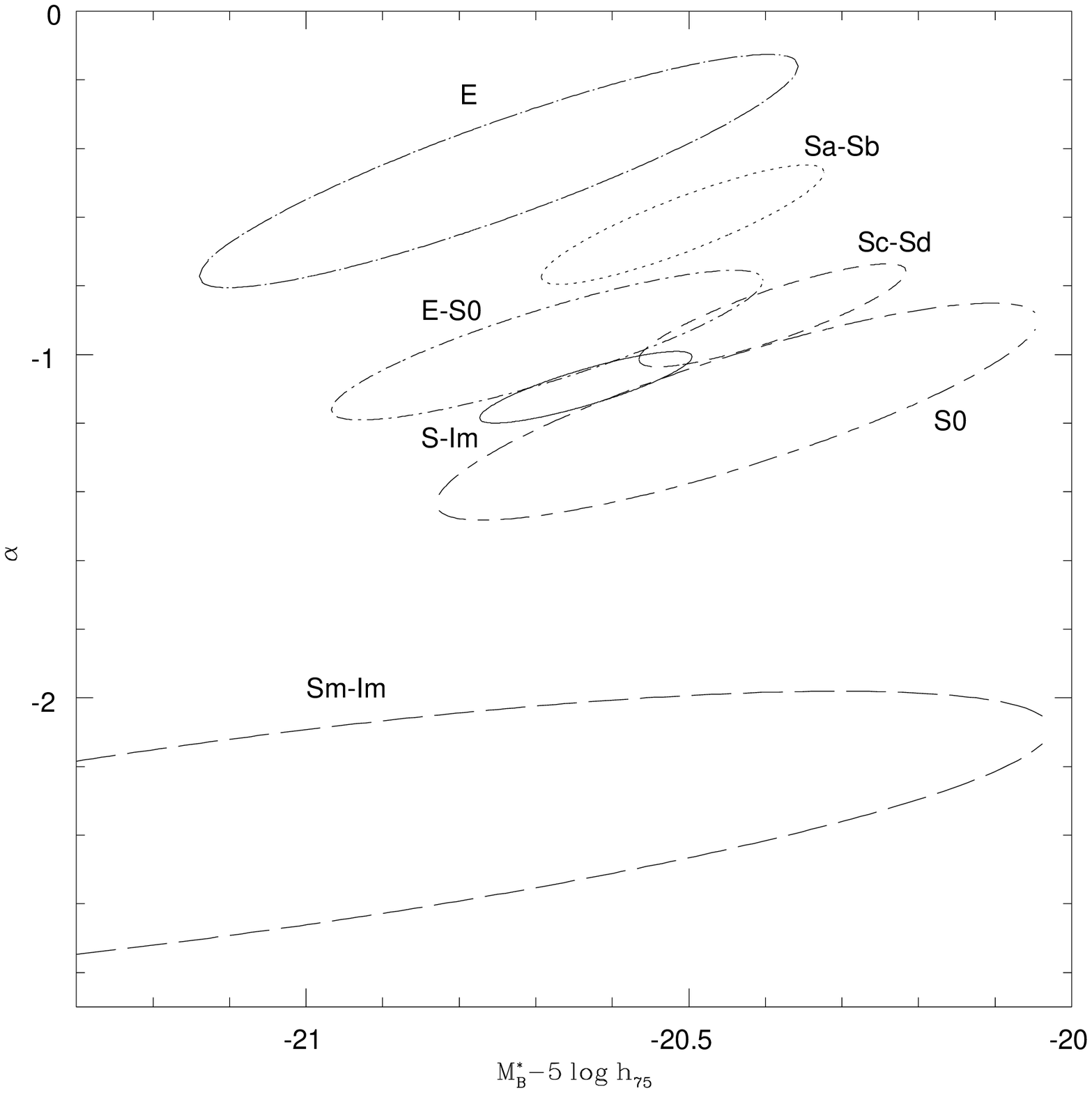}
$\ \ \ \ \ \ $\\
\vspace{18.0truecm}
$\ \ \ $\\
%\vspace{-20mm}
{\small\parindent=3.5mm {Fig.}~4.---
We   plot contours  representing the 1  $\sigma$
deviation  from the best-fitting  values  of $\alpha$ and  $M_{B}^{*}$
relative   to the  LFs   shown  in  Fig. 3 for   different
morphological types.}
\vspace{5mm}
%\begin{multicols}{2}

\clearpage

\begin{center}
{\em 3.3 The luminosity function of group galaxies}
\end{center}
 
With the purpose  of investigating  on the  effect of  the presence of
galaxy systems  on the LF  determination, we recalculate the LFs after
removing the members of the systems containing more than 5, 10, and 20
galaxies (e.g., removing 114, 42, and 10  systems for a total of 1437,
830, and 355  members, respectively, in the  specific case of the Mark
III multi-attractor model).  We  never find any significant difference
in  $\alpha$ and  $M_B^{*}$ with  respect to the   values tabulated in
Table 1 for the whole sample.

Furthermore, we calculate the LFs for galaxy members of groups and for
field (non-- grouped) galaxies,   separately, according to  the various
distance models. Table 3 lists  the results.  We find
no significant differences    between  the LF  derived  from   the two
subsamples, whenever, for constructing the sample of grouped galaxies,
we take all systems or a subsample of systems dominated by groups with
few members (e.g., all the systems with  $n>5$ members).  For all sets
of distance models, we detect some marginal systematic effects (at the
$\sim1\sigma$  confidence  level)  if we  restrict   ourselves to  the
richest systems containing $n\ge10$, $n\ge15$,  $n\ge20$ members.

%%   macro per inserire tabelle
%%TAB3
\vspace{14mm} 
\hspace{-15mm}
\begin{minipage}{18.5cm}
\renewcommand{\arraystretch}{1.25}
\renewcommand{\tabcolsep}{3.2mm}
\begin{center}  
\vspace{-3mm}
TABLE 3\\
\vspace{2mm}
{\sc The Schechter LFs for field and group galaxies.\\}
\footnotesize
\vspace{2mm}
\input{tab3}

\end{center}
\vspace{3mm}
\label{lfdatagroups}
\end{minipage}

\clearpage

%%%%%%%%%%%%%%%%%%%%%%%%%%%%%%
%%   macro per inserire tabelle
%%TAB4
\vspace{10mm} 
\hspace{-15mm}
\begin{minipage}{18cm}
\renewcommand{\arraystretch}{1.9}
\renewcommand{\tabcolsep}{0.6mm}
\begin{center}  
\vspace{-3mm}
TABLE 4\\
\vspace{2mm}
{\sc The parameters of the LFs from various samples.\\}
\scriptsize
\vspace{2mm}
\input{tab4}

\end{center}
\vspace{3mm}
\label{lfliterature}
\end{minipage}

 The
LF of their members tend to show somewhat brighter values of $M_B^{*}$
with respect to the field value, with no systematic shift of $\alpha$.
Although  the average number  of members of  a group tends to decrease
with increasing distance,  the tendency  is not  so  strong to  affect
significantly the above--mentioned results.

\begin{center}
{\bf 4 Discussion}

{\em 4.1 Comparison with previous total galaxy luminosity functions}
\end{center}

In Table 4 we show  the Schechter function parameters
of  some recent redshift  surveys   of field galaxies.  All  published
values have  been reduced to $H_0=75\;km\;s^{-1}\;Mpc^{-1}$.  We quote
the original photometric passband of the  survey, but we transform the
characteristic absolute magnitude $M^{*}$  to $M_B^{*}$ through simple
magnitude  offsets; namely $bj  -  r  = +   1.1$,  which is  the  mean
rest--frame  color of the Las  Campanas redshif survey (LCRS) galaxies
(Lin  et al. 1996), $bj -  R = +1.3$,  which is  the median rest-frame
color of the  Century  Survey (CS)  galaxies  (e.g., Buta  \& Williams
1995; Geller et al.  1997), $bj - m_Z =  -0.45$ (the shift of the blue
Zwicky   magnitudes  $m_Z$   comes  from the    comparison  of  galaxy
number--magnitude counts in Shanks  et al.  1984);  $B - bj = -0.3$ as
given by Efstathiou et al. 1988; $B - B(0) = -0.3$  (the zero point of
the scale of the $B(0)$ magnitudes agrees with the $bj$ one within 0.2
mag, according to da Costa et  al.  1994). The  formal errors given in
Table 4 do not include uncertainties related to color
transformations. In the following, we  quote the results given in the
literature as scaled into the color and $H_0$ we are using.

\clearpage

%%FIGURE 5%%%
%\end{multicols}
%\begin{figure}
\includegraphics{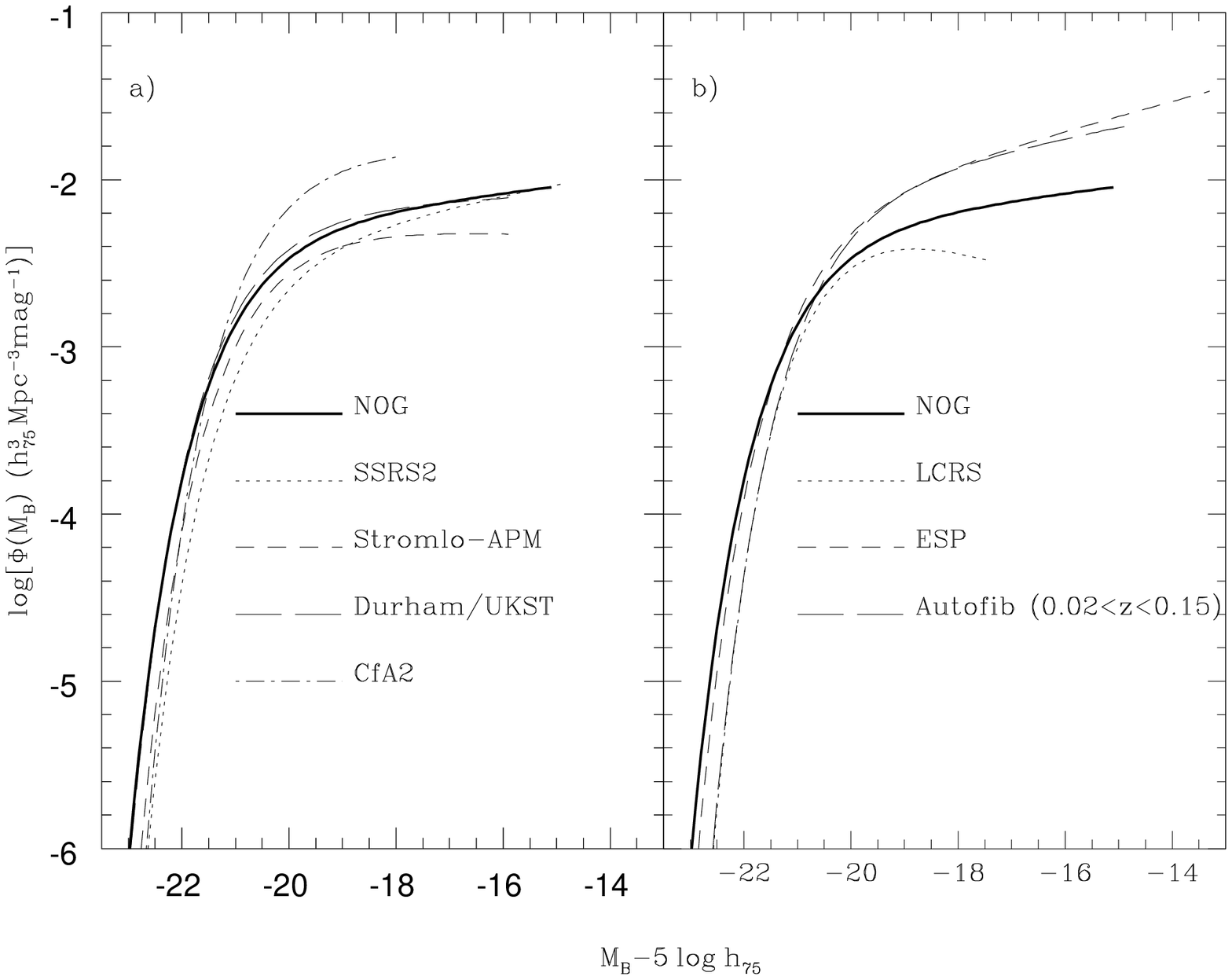}
$\ \ \ \ \ \ $\\
\vspace{14.0truecm}
$\ \ \ $\\
%\vspace{-20mm}
{\small\parindent=3.5mm {Fig.}~5.---
This figure shows  a comparison between  the NOG
LF we calculate in the case of the multi-attractor model based on Mark
III data,  and previous  LFs obtained  in  the literature  from recent
redshift   surveys   of  low     depth (a)  and    intermediate  depth
(b).}
\vspace{5mm}
%\begin{multicols}{2}

Fig. 5 illustrates the comparison of our normalized
LF (relative    to the case of  Mark   III-based distances) with those
obtained   from  some   low-depth  (Fig. 5a)  and
intermediate-depth (Fig. 5b) redshift surveys   of
field galaxies. 

An accurate comparison  between the LFs is  impeded  by the fact  that
applying   simple  constant  magnitude   offsets   does  neglect  more
complicated effects,  which   are due for   example  to galaxy colors,
morphologies, and surface brightnesses.   However, among the LFs there
are    certainly meaningful  differences  which are    due to problems
mentioned  in \S 2.  In  any  case, in  the optical  and near-infrared
surveys  mentioned  in  this  section  there   is  no indication  of a
systematic  flattening   of   the   slope $\alpha$,    a  simultaneous
brightening     of $M_B^{*}$, and    a    progressive decrease  of the
normalization factor  with  brighter surface brightness  limits of the
survey, as  expected in   the  case in which   observational selection
effects    depending on brightness  surface  limits  were the dominant
effects (e.g., Dalcanton, Spergel \& Summers 1997b). 

The  $M_B^{*}$-values given in   Table 4   are always
appreciably fainter   (by $\sim$0.1--0.6 mag)  than our  values, which
give  a   slower  decline of the   number  of  luminous  galaxies with
increasing luminosity.

Our $\alpha$--values well agree with the average results obtained from
the  shallow  surveys, such as    the  Stromlo-APM, CfA2,  SSRS2,  CS,
Durham/UKST surveys,  and  from  intermediate-depth  redshift surveys,
such as the ESP and Autofib survey (restricted to $0.02<z<0.15$).  The
only  very discrepant value of $\alpha$  is that coming from LCRS. Its
declining  faint--end slope ($\alpha\sim$-0.7) can not   be due to the
fact   that  galaxies are selected    in  the  red  band, because  the
R-selected CS sample  yields a much  steeper  slope. It can be  partly
related to the fact that  a Schechter form does not  fit very well the
LCRS  data around $M_B\sim$-19.5 and in  the faint end (see Bromley et
al.  1998)  and mostly to  the  fact that LCRS  may be
biased against
observing   low-luminosity galaxies, because   of the survey selection
criteria (Geller et al. 1997). 
 
In the  intermediate  luminosity interval ($M_B\sim$-20),  our results
point   to   a    fairly   low     amplitude  for   the      LF   (see
Fig. 5), in  substantial agreement with the results
coming from   several  shallow redshift surveys,   especially with the
Durham/UKST  survey,  which  has yielded   the  LF which  most closely
resembles our results.

It is interesting, albeit less easy, to compare our results with those
coming  fron the ORS. As  a matter of fact,  as warned  by Santiago et
al. (1996), it is difficult to properly compare the ORS LFs with other
published  LFs in the literature (including  our results), because the
ORS LFs are not corrected for Malmquist bias. Fig. 6 shows
the  comparison   of our normalized   LF   (relative to  the  case  of
Hubble-flow distances  in the  LG   frame) with the ORS  LFs  directly
obtained from  the UGC and ESO  subsamples by Santiago  et al. (1996),
who  parameterized  the   ORS  LFs through  a   generalization of  the
Schechter  form.  For this   comparison we  apply  no magnitude offset
(since the ESO  B system does not  deviate too much from  the standard
RC3 B system; see e.g., Paturel,  Bottinelli \& Gouguenheim 1994), and
no correction for differences  in the mean  galaxy density between the
UGC   and ESO subsamples.   The NOG LF  appears  to be in satisfactory
agreement  with the  ORS LFs  at   intermediate and faint magnitudes,
whereas the bright end  of the former  seems to be shifted brightwards
(like in the case  of Fig. 6), suggesting magnitude
offsets of $\sim$0.5 and $\sim$0.7 mag, for the ESO and UGC subsamples
respectively. Notably, the   latter  offset (which refers  to   Zwicky
magnitudes) is  much smaller   than  the amount  ($\sim$1.2-1.3 mag  )
required to match the NOG LF and the bright  end of the CfA2 LF (which
also uses Zwicky photometry). 

\clearpage

%%FIGURE 6%%%
%\end{multicols}
%\begin{figure}
\includegraphics{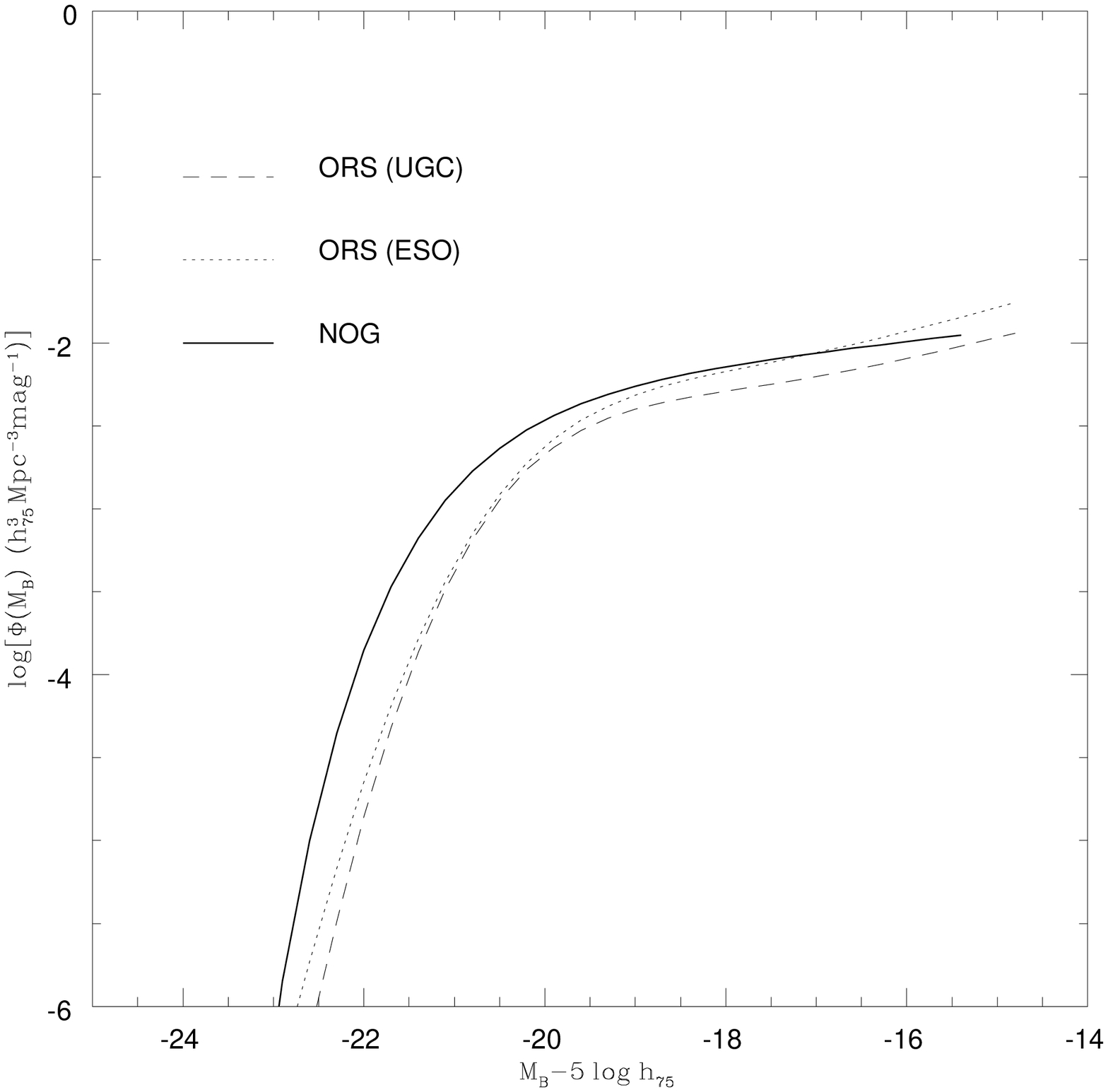}
$\ \ \ \ \ \ $\\
\vspace{14.0truecm}
$\ \ \ $\\
%\vspace{-20mm}
{\small\parindent=3.5mm {Fig.}~6.---
 This figure shows  a comparison between the NOG
LF that we calculate in the  case of redshift-distances obtained using
the Hubble relation in the Local Group frame  and the ORS LFs obtained
directly from the ESO and UGC subsamples of the ORS.}
\vspace{5mm}
%\begin{multicols}{2}

The anomalously high  value of the CfA2  normalization is likely to be
due   to strong local inhomogeneities (e.g.,   the "Great Wall").  The
quoted CS normalization, being   the  average value for  the  velocity
range  between 18 000  and 28 000 \ks,  is also high,  because in this
range  this survey slices  through a  portion  of the  Corona Borealis
supercluster. Remarkably, both  CfA2  and CS evidence that  $\phi^{*}$
increases by a  factor $\sim$2 from  nearby regions ($cz<5000$ \ks) to
distant regions   ($cz\sim10 000$ \ks    and $\sim25 000$  \ks),  thus
admitting a normalization  consistent with  our  value in the   nearby
regions. 

The deeper redshift  surveys  (ESP,  LCRS,  Autofib)  disagree in  the
amplitude of the LF at intermediate and faint magnitudes.  Our results
are intermediate   with respect to the low   LCRS value and   the high
values of the ESP and Autofib samples.  Zucca  et al.  (1997) stressed
the possibility of reconciling the high ESP normalization with the low
values given   by  other surveys,  by noting   that  the nearby region
($cz<14 000$  \ks)  encompassed by the ESP   is underdense and  that a
recalculation of  the LF for a subsample  of  bright ESP objects would
gave results    consistent with those    derived from the Stromlo--APM
survey.   This view   seems  to  be   qualitatively supported by   the
preliminary results of  the  intermediate--depth 2dF redshift   survey
(whose galaxies are selected from  the same catalog  used in the  last
survey);  also  this  LF  hints at  a greater   normalization than the
Stromlo--APM one (Maddox 1998). 

Noticeably, also the small--field  deep near-infrared by Glazebrook et
al.   (1995b)  and  Cowie  et al.   (1996)  disagree  in the K-band LF
normalization. The normalization found from  the latter, larger survey
is close to that coming from  shallow near-infrared surveys (Mobasher,
Sharples \& Ellis 1993; Gardner et al.  1997; Szokoly et al. 1998) and
is few times smaller than that of the former survey.

From  the  shallow IRAS 1.2  Jy survey  conflicting results about the
presence  of a local void  have emerged ({\it   cf} Koranyi \& Strauss
1997 and Springel \& White  1998), whereas the deeper QDOT-IRAS survey
suggests a  moderate local   void   extending out to   $\sim$9000  \ks
(Saunders et   al.  1990,  Fig.  11).   More reliable results   may be
expected from the  larger PSCz survey  (e.g., Canavezes  et al.  1998)
which is under study.

To sum up, some optical redshift surveys  (CfA2, CS, ESP) favor a high
average normalization,   but   at the   same  time  hint  at  a galaxy
underdensity in  the nearby regions. If  we rely on the results coming
from these   surveys, we should conclude  that   the local universe is
really underdense.  But optical and IRAS  redshift surveys disagree on
the  extent,  depth, and even  real existence  of a local underdensity
(see \S 4.4 for further discussions on this point). 

\bigskip
\begin{center}
{\em 4.2 Comparison with previous type-specific galaxy luminosity 
functions} 
\end{center}

\bigskip

Our  morphological-type dependence of the  slope of the optical galaxy
LF is in substantial agreement with the results  obtained by Marzke et
al.  (1994b, 1998) from the CfA2 and the  SSRS2 samples. The number of
Sm--Im objects rises towards the faint  end even steeper in our sample
than  in those two samples   (where $\alpha\sim$-1.9).  Moreover,  the
difference between the   $\alpha$-values  of E  and S0  galaxies  (and
correspondingly the  shortage   of low-luminosity E  systems)  is more
marked in our  sample than in the  CfA2 sample. The difference we find
is fully   consistent with  claims that in   the  field the transition
region between giants and dwarfs is scarcely populated  in the case of
ellipticals and better populated  in the case of lenticulars,  because
galaxy bulges tend  to weaken with  decreasing luminosities and become
dominated by disk or  lens-like components (see,  e.g., the RSA survey
of dwarf galaxies by Vader \& Chaboyer 1994). 

Nevertheless,  the dependence of  the  slope on the morphological type
across the whole Hubble  morphological sequence is  not so striking in
our  sample as  in  the  Stromlo-APM  sample,   where  the very  steep
declining  of   early--type galaxies  at the  faint   end (Loveday  et
al. 1992) can be ascribed to the incompleteness of early types at high
redshift and/or faint magnitudes (as discussed by Marzke et al. 1994b,
and Zucca, Pozzetti \& Zamorani 1994). 

Our analysis establishes that the $M_B^{*}$-value of E galaxies is not
unusually bright, supporting Marzke  et  al.'s (1994b) guess  that  an
uncorrected  dependence of  the    Zwicky magnitude scale on   surface
brightness  (e.g.,  Ichikawa \&  Fukugita 1992)   could have bias  the
magnitudes of the CfA2 ellipticals. 

At  middle  and faint magnitudes  ($M_B>$-20),  our  spiral LF  has an
amplitude closer to those of the CfA2 and SSRS2 spiral LFs than to the
lower one of the Stromlo-APM late-type LF. At the same magnitudes, the
amplitude of our E-S0 LF is close to that of the SSRS2 E-S0 LF. 

Interestingly, the spectral type-dependence  of the  galaxy luminosity
function as derived from the LCRS ( Bromley et al.  1998 a, b) and the
Autofib redshift survey (Heyl  et al. 1997)  shows a  more progressive
steepening of   the slope  from early  types  to  late types than  the
morphological-type  dependence derived in this  paper.  This is due to
the    fact  that galaxy  typing  based   on the visible morphological
appearance is not  very tightly correlated with  galaxy classification
based on spectral schemes (e.g., Kennicutt 1992; Connolly et al. 1995;
Zaritsky, Zabludoff \& Willick 1995; Katgert et  al. 1999) because the
latter method   is  very sensitive  to the   degree  of star formation
activity. 

Our  study  lends further support to   the  fact that  blue, gas-rich,
late-type spirals and irregulars constitute a steeply rising component
of the field galaxy LF. Since these galaxy types dominate in the field
at very low luminosities (e.g., Ferguson  \& Binggeli 1994; Sprayberry
et al. 1997), our finding leads us to  envisage an upturn of the field
LF in the  faint end  (though not  brightwards $M_B\sim$-15). This  is
consistent with the growing evidence that the field galaxy LF shows an
upturn   from  the    Schechter  form  at    faint   magnitudes (where
magnitude-limited  samples   become  severely   censored    by surface
brightness selection  effects).  There is straightforward evidence for
this faint--end steepening in the CfA2 and ESP data (through an excess
of low-luminosity galaxies  above the extrapolated Schechter function)
and in the counts of the statistical excess of faint APM galaxies seen
in projection around   Stromlo--APM redshift survey  galaxies (Loveday
1997). 

\clearpage

\begin{center}
{\em 4.3 Comparison with previous luminosity functions of group and 
cluster galaxies} 
\end{center}

The most  recent and  extensive  analyses  of the cluster   galaxy  LF
(Lumsden et al. 1997; Valotto et al.   1997; Gaidos 1997; Rauzy, Adami
\& Mazure 1998) concur to $M_B^{*}$-values brighter by a few tenths of
magnitudes   than those of  the field  galaxy LFs  mentioned in \S 4.1
(including our LF and the LFs relying on the same photometric system),
whereas there  is  no agreement on  whether the  LF slope  $\alpha$ of
members of (poor and/or rich) clusters significantly differs than that
of  field   galaxy LF  (see the conflicting   results reached   by the
above--mentioned authors and also by L\'opez--Cruz et al. 1997). 

Notably, that tendency is consistent with the environmental dependence
of the LF we recognize in \S  3.3. We find  that this dependence is so
much weak  (at least in  the  systems of moderate  richness which  are
typical  for our sample) that the  presence of galaxy systems does not
affect  significantly the field  galaxy LF.  This  is at variance with
some   pronounced cluster effects   (i.e.,  a steepening of the  slope
$\alpha$   in samples containing clusters)   reported  by Marzke \& da
Costa (1997) for a small subsample drawn from the SSRS2. 

Our results are in line with the claims of some luminosity segregation
with density for several  galaxy samples (e.g.,  Park et al. 1994, for
the CfA2 galaxies; Giuricin  et al.  1995,  for the Local Supercluster
galaxies; Loveday et al.  1995 for  the Stromlo-APM galaxies, Willmer,
da Costa \& Pellegrini 1998  for the SSRS2 galaxies). This segregation
was found to be weaker than  morphological segregation with density by
Iovino et al. (1993) for  the Perseus-Pisces supercluster galaxies and
by Dom\'inguez-Tenreiro, G\'omez--Flechoso   \& Mart\'inez (1994)  for
the CfA1 galaxies,  whereas   it was considered significant  only  for
early-type luminous galaxies by Hasegawa \& Umemura (1993). 

Among recent  works in  which  environmental  effects  on  the  LF are
searched for through  a  specific comparison between group   and field
galaxy LFs,  an approach similar  to ours  was followed  by Ramella et
al. (1999) and Bromley et al.  (1988a,b), who divided their respective
ESP  and LCRS  population into  field  galaxies and  galaxy members of
groups (or  rich groups), identified  by means of  the {\it friends of
friends} algorithm. 
 
The former  authors found that, compared to  the LF of field galaxies,
the LF of group galaxies exhibits a  brighter $M^{*}$ (by a few tenths
of magnitude), which brightens with the increasing richness of groups,
whereas  the slope $\alpha$  does not  change  significantly  from the
field to the  groups. Ramella et  al.'s  (1999) results on changes  of
$M^{*}$ are quantitatively  compatible with our findings, although the
authors claimed a larger (2 $\sigma$) statistical significance. 

On the other hand, the latter authors detected  a steepening of the LF
slope  for  galaxies located  in regions of  higher density.   This is
caused by a strong environmental effect (in the same  sense) on the LF
of early-type galaxies (there is no effect on late types).  We must be
wary  that their steepness-density   relation  (which we do not  find)
could reflect some property  of the survey  (which shows a north-south
discrepancy  in total and   type-dependent  LFs) rather  than  a  true
environmental effect.

\begin{center}
{\em 4.4 Relation of the galaxy luminosity function to galaxy counts}
\end{center}

The  following arguments bearing on the  observed galaxy number counts
provide a  better evidence for an underdensity  of the nearby universe
than those presented in \S 4.1. As discussed by several authors (e.g.,
Driver \& Phillipps 1996,  and the reviews  by Shanks 1989, and  Ellis
1997), in  the  absence   of   appreciable galaxy  evolution,   a  low
normalization   of   the local  galaxy  LF,    as  derived  by us,  is
inconsistent with the high counts which are generally observed already
at intermediate optical and near-infrared magnitudes (i.e., roughly at
$18<B<20$ mag, $13<K<15$ mag). The observed counts at these magnitudes
would imply a LF normalization greater by a factor $\sim$1.5 than that
of our LF. A high normalization  is customarily assumed in most models
of  faint  galaxy counts,  which  are   generally normalized   to  the
observations in  the  magnitude  intervals quoted above    rather than
locally. Workers (e.g., Koo, Gronwall \& Bruzual 1993; Gronwall \& Koo
1995), who claimed  that the high bright-end  counts can be reasonably
fitted by no-evolution or mild evolution models, adopted the view that
the local LF is ill-defined. 
 
We deem  it more reasonable to explain  the  underprediction of galaxy
counts   by   conceiving a local   underdensity   rather than  a rapid
evolution (Maddox et al. 1990) of the bulk of the galaxy population at
the  corresponding   low redshifts  of  $z\sim$0.1  -- 0.2  because on
theoretical grounds little  galaxy evolution is  expected up to  these
redshifts  and because very deep   redshift  surveys which have  begun
measuring the  galaxy LF  out  to $z\sim$0.5 --  1 (e.g.,  Ellis 1997)
reveal little evolution in the LF of most  of galaxies, except for the
late-type low-luminosity population. 

Consistently, no--evolution models based on our local LF  , but with a
normalization  roughly  1.5   times our value,   would  satisfactorily
predict the   counts  of  E/S0   and Sabc  spirals,  but   would still
underpredict   those of late-type   spirals   and irregulars, for  the
morphologically classified sample of the faint field galaxies from the
HST     Medium Deep Survey (Driver,   Windhorst    \& Griffiths  1995;
Glazebrook et al. 1995a; Abraham et al.  1996) and the deep HST fields
(Odewahn et al. 1996). 

As  for  the compatibility  of   the observed bright  end counts  with
no-evolution     or passive  evolution    models, recent  observations
disclaimed the very steep   slope of the   APM counts reported in  the
$16<bj<19$ mag range  by Maddox et al.  (1990) --- who explained their
counts by invoking a large amount of evolution at low redshift --- and
attributed the steepness of the APM counts to systematic errors in the
non-linear APM  photometry.  The same effect  may  account for the low
counts of Heydon-Dumbleton, Collins  \& MacGillivray (1989)  and Zucca
et al. (1997) at   bright COSMOS-$bj$ magnitudes.  The flatter  slopes
which fit   the   bright-end blue  counts   of  recent  studies (Weir,
Djorgovski \& Fayyad 1995;  Gardner et al.   1996, Huang et al.  1997,
Bertin \& Dennefeld 1997) do not largely  depart from the expectations
of no-evolution models, but always require a higher normalization than
our local one. 

A more  convincing evidence   for  such a   departure comes  from  the
bright-end  K-band number counts.  At bright  magnitudes in the K-band
($10<K<17$ mag),  where the K-correction  is known better than  in the
optical  band and is only a  weak function of  morphological type, the
number counts delineate  a slope which  is too steep to  be consistent
with simple no-evolution models (Huang et  al. 1997).  Detailed models
of these counts have explained  this effect through  a large region of
local  underdensity of  galaxies (by  a   factor $1.7  -- 2.4$), which
extends  out  to $z\sim$0.1  -- 0.2,  in the case  of no-evolution, or
through a combination of some  local underdensity and a mild evolution
(Phillips \& Turner 1998). 
 
The initial  steep rise of the radio  source counts between  fluxes of
$\sim$10  Jy and $\sim$1  Jy (in  the  1.4 --8.4 GHz  frequency range)
(e.g., Windhorst et al. 1993) also may suggest the presence of a local
underdensity of  nearby radiogalaxies, since their strong cosmological
evolution is  thought  not to  start  until $z\sim$0.3   (e.g., Condon
1989). 

The  existence of a large-scale local  underdensity implies that local
measurements   underestimate (overestimate)    the  true   value    of
$\Omega_{0}$ ($H_{0}$).  The  local value of  $H_0$ is  expected to be
higher than the global value (Turner, Cen \& Ostriker 1992) by as much
as $\sim$20\%  for $\Omega_{0}=$1 and $\sim$10\% for $\Omega_{0}=0.2$,
in  the case  of  an underdensity  by a factor  of  1.5 (see also Shi,
Widrow \& Dursi 1996).  A moderate density  contrast (of this order of
magnitude   or  smaller) is   lower than  the   value  favored by  the
above-mentioned  models of   bright-end K-band  galaxy  number  counts
(Phillips  \&  Turner  1998),   but it is    more consistent  with the
following    observational  constraints    relative  to   Hubble  flow
perturbations.  As a  matter of fact, from  the peculiar velocities of
supernovae Zehavi et  al. (1998) found  a $\sim$7\% deviation from the
Hubble  law consistent     with  a void   of  $\sim$20\%  underdensity
surrounded by a  dense wall at 7000  \ks (which roughly coincides with
the local  Great Walls).  This  is compatible  with the  corresponding
upper limit ($\sim$7\%)   obtained from the  brightest cluster  galaxy
Hubble diagram    (Lauer   \&  Postman 1992;    Lauer  et  al.  1998).
Differences between  the local and global values  of $H_0$ as large as
$\sim$10\%  on scales of $\sim$30,000   \ks are still fully compatible
with the  limits sets  by the  measurements  of cosmological distances
through high redshift supernovae (e.g., Kim et  al.  1997), but become
only marginally compatible  with  the limits  imposed  by the  current
knowledge  of cosmic  microwave background  anisotropies  (e.g., Wang,
Spergel \& Turner 1998). 

\begin{center} 
{\bf 5 The selection function}
\end{center}

The NOG  sample,  being magnitude--limited, becomes sparser  at larger
distances  due to  the  increasing  loss of  galaxies  caused  by  the
apparent magnitude cutoff. This effect  is quantified by the selection
function $S(r)$  which expresses the  fraction   of galaxies that  are
expected to satisfy the sample's selection criterion. The incompletion
factor $F(r)$, which expresses the number of galaxies that should have
been cataloged   for  each object present  in  the  sample  at a given
distance $r$, is related to $S(r)$ by the expression $F(r)=1/S(r)$. 

The selection function is given by 
\begin{equation} 
S(r)=f      \cdot      \left\{\begin{array}{lr}    \frac{\displaystyle
\int_{max[L_{s},L_{min}(r)]}^{\infty}\;\;\Phi_{o}(L)dL} {\displaystyle
\int_{L_{s}}^{\infty} \;\Phi_{o}(L)dL}  &  \textrm{ if $r > r_s$}\\  1 &
\textrm{ if $r<r_s$} \end{array} \right. 
\label{selfun}
\end{equation}

\noindent where $\Phi_o$ is the convolved Schechter LF, $f=0.8$ is the
average  completeness  level,  $L_{min}(r)$ is the  minimum luminosity
necessary  for a galaxy at  distance $r$ (in  Mpc) to make it into the
sample,  and    the  integral is   cut  off   at the  lower   limit of
$L_s=L_{min}(r_{s})$ corresponding to $M_s=  -15.12  + 5 \log  h_{75}$
(see \S 2.2).  Through Monte Carlo simulations Santiago et al.  (1996)

\noindent demonstrated that an 
unbiased estimate of the galaxy density field can
be obtained in the presence of random errors on magnitudes if one uses
the convolved LF in eq. \ref{selfun}. 

Fig. 7  shows the incompletion  factor $F$ as a function of
the  distance $r$ (in \ks) for  different  velocity field models.  The
incompletion factor $F$ is always  a smaller in the  case of the  Mark
III multi-attractor model  than in  the case   of the cluster   dipole
model, but $F$ appears to be little sensitive to the adopted models.

The dependence of the selection function on redshift-space distortions
has been  addressed analytically by  Hamilton (1998)  in the  case  of
linear theory. His work is relevant in order to understand the results
given in Fig. 7, and then it is mentioned here. 

Denoting the  redshift-space    coordinate  by $s$,    the  real-space
($F^{r}(r)$)  and redshift-space ($F^s(s)$)  incompletion functions in
the LG frame are related as follows (see eq. (4.85) in Hamilton 1998): 

\begin{equation} \ln \left(\frac{F^r(r)}{F^s(s)}\right) =
\int_r^{r_{max}}      \frac{d^2 \ln    F^s(r')}{d\ln   s'^2}    \left(
\left\langle\frac{v}{r}\right\rangle_{r'}  -  \left\langle  \frac{{\bf
\hat{r}}}{r}      \right\rangle_{r'}\cdot      {\bf     v}^{LG}\right)
\frac{dr'}{r'} \label{eq:hamilton} 
\end{equation}

\noindent where   $r_{max}$ is the  depth  of  the  sample.  Note that
averages   are not  ensemble averages, in   which case  the real-  and
redshift-space  $F$ would be equal, but  are performed over the actual
galaxy sample. 

\clearpage

%%FIGURE 7%%%
%\end{multicols}
%\begin{figure}
\includegraphics{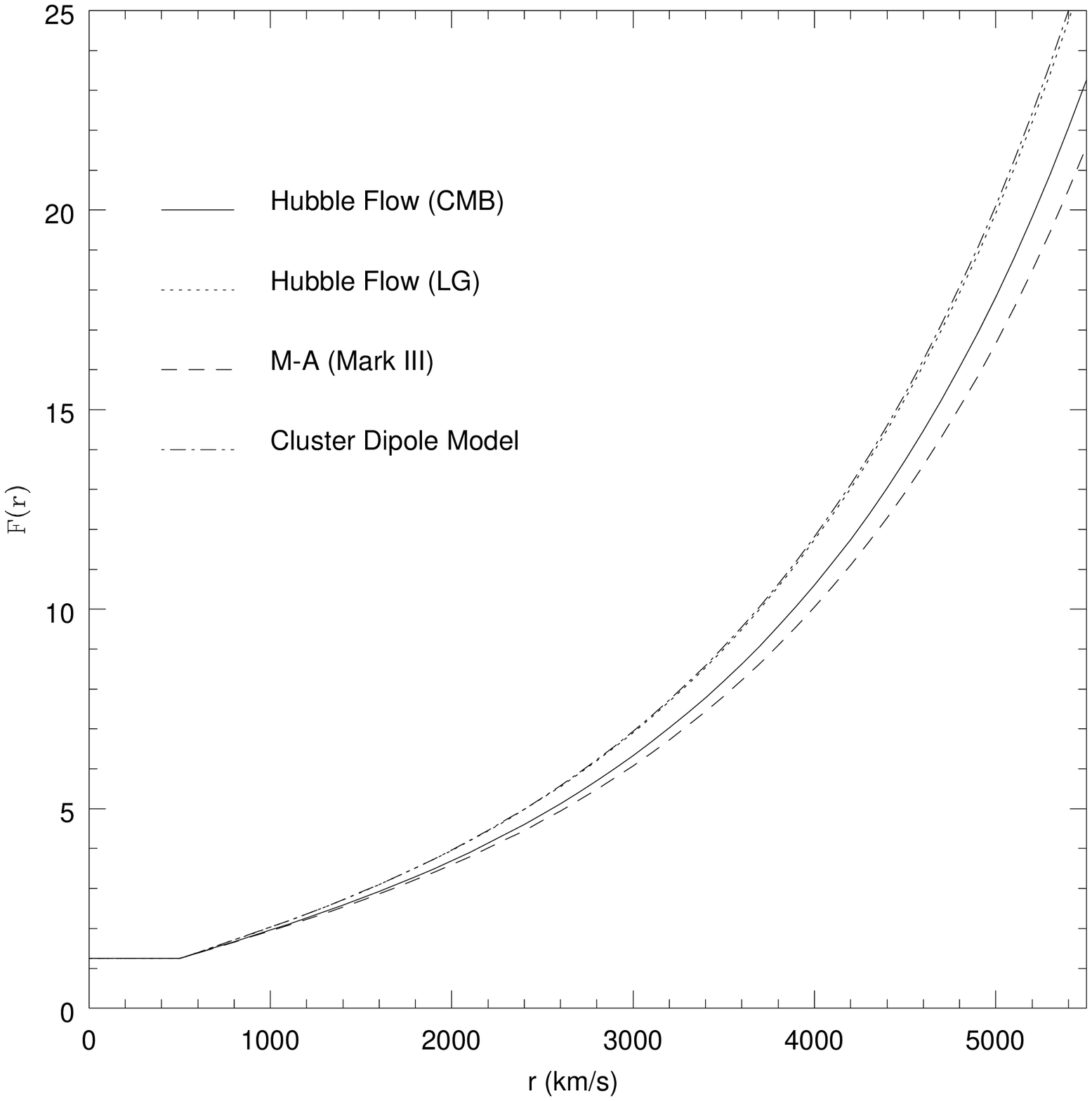}
$\ \ \ \ \ \ $\\
\vspace{17truecm}
$\ \ \ $\\
%\vspace{-20mm}
{\small\parindent=3.5mm {Fig.}~7.---
We plot  the   incompletion  factor  $F$  as   a
function  of galaxy distance  $r$  (expressed in  km/s) for  different
peculiar velocity field models.} 
\vspace{5mm}
%\begin{multicols}{2}

The second term inside  the parenthesis of the  right hand side of eq.
\ref{eq:hamilton}   averages to  zero   for a full-sky   catalog; this
highlights a further  advantage of the  full-sky coverage.   The first
term in the same parenthesis is the average, at the  depth $r$, of the
line-of-sight   peculiar velocity  $v$   over $r$.   Then, the largest
contribution  to the differences between  the real- and redshift-space
incompletion functions  comes  from  the  very nearby  universe,  both
because of  the  $1/r$ term  and   because the  velocity $v$  is  less
affected by local fluctuations if the average volume is larger. 

With  eq \ref{eq:hamilton} ,    considering that   the     logarithmic
derivative is positive for common flux-limited samples, it is possible
to  understand the behaviors of the  different models. As  a matter of
fact, any model which gives  a local overdensity has negative $\langle
v/r \rangle_{r}$ (converging   velocities). This is  the case  for the
Mark III multi-attractor model,  which   has a nearby attractor   (the
Virgo cluster) and  two other attractors (the  Great Attractor and the
Perseus-Pisces supercluster) at distances of $\sim$4000 -- 5000 km/sec
(the distant Shapley concentration gives a significant contribution to
the LG motion,  but covers  a small solid  angle).   As a consequence,
galaxies are brighter (more distant) on average in real space, and the
corresponding incompletion factor is consequently smaller. 

On the  other hand,  the cluster dipole  model  gives a much  smoother
version of the  local velocity  field.  As  a  consequence, the  local
overdensity  is low and  the incompletion  function remains similar to
the LG one.

\begin{center}
{\bf 6 Summary and conclusions}
\end{center}

We evaluate the field galaxy LF and  the related selection function of
the    nearly    complete,    magnitude-limited (B$\leq14$  mag)   and
volume-limited ($cz<$5500   \ks) NOG (Nearby Optical   Galaxy) sample.
Our derivation of the LF is performed using the locations of field and
grouped galaxies in real distance space, since we adopt sets of galaxy
distances based on different peculiar velocity field models.  Our main
results can be summarized as follows: 

i) The local field galaxy LF is well described by a Schechter function
with         $\alpha\sim$-1.1,          $\Phi^{*}\sim0.006\;Mpc^{-3}$,
$M_B^{*}\sim$-20.6       down     to    $M_B=-15.1$    ($H_0=     75\;
km\;s^{-1}\;Mpc^{-1}$). The exact values  of the Schechter  parameters
little depend   on the adopted  velocity field  models, since peculiar
motion effects are  of the order of statistical  errors (at most, they
cause variations of 0.08 ($\sim1\sigma$ error) in $\alpha$ and 0.2 mag
($\sim2\sigma$ error) in $M_B^{*}$).  Thus, the local field galaxy LF,
as well as    the intimately related selection   function  of the  NOG
sample, are little sensitive to peculiar motions effects. On the other
hand,  the  details of peculiar   velocity models  have  a quite large
impact on the local galaxy density on the smallest scales. 

ii) Compared to  previous local     field  galaxy  LFs, our LF      is
characterized  by a normal  (moderately   steep) slope $\alpha$ and  a
$M_B^{*}$-value  which is, on  average, brighter  by  0.4 mag than the
norm.  The last result is explained from the fact that the photometric
measures of the  NOG galaxies, being total  B magnitudes corrected for
Galactic absorption,  K-dimming,    and  internal  absorption,  better
represent the true galaxy light. 
 
iii)  Our morphological-type  dependence of  the galaxy LF  is in much
closer  agreement with the results   obtained from the CfA2 and  SSRS2
samples (Marzke et  al. 1994b, 1998) than  with those derived from the
Stromlo-APM sample (Loveday et al. 1992). The dependence of the galaxy
LF from the morphological type appreciably differs from its dependence
on the spectral classification as given in the literature. 

The LF shape  of E-S0 galaxies does not  differ significantly from the
that of spirals and  irregulars.  On the other   hand, the E  galaxies
clearly decrease in  number  towards  low  luminosities,  whereas  the
number of   late-type spirals and   irregulars show a very  steep rise
towards the faint end (with $\alpha\sim$-2.3 -- -2.4).  This behaviour
suggests a   similarly fast increase  in  the number of low-luminosity
galaxies in the faint end unexplored in the NOG sample (at $M_B>-15$). 

iv) The presence of galaxy  systems in the  NOG sample does not affect
significantly the shape of  the the field  galaxy LF.  As a  matter of
fact, environmental effects on the total  LF are proved to be marginal
in the NOG sample. However, the total LF of  the galaxy members of the
richest systems  tends to show a somewhat  brighter value of $M_B^{*}$
(at the 1$\sigma$ level), which is in line with several earlier claims
of galaxy   luminosity segregation with   density (in  particular with
Ramella et al.'s (1999) results). 

This   effect is unrelated to   an  increased proportion of early-type
galaxies in   galaxy systems. A  luminosity  segregation  with density
could  be related to  the processes of  biased galaxy formation (e.g.,
Kaiser  1984; Rees 1985) and/or could  be induced by late evolutionary
processes ---  such as tidal shocks  (e.g., Nogushi \& Ishibashi 1986;
Sanders  et al. 1988), ram  pressure effects  (Dressler \& Gunn 1983),
and  merging effects  (e.g., Barnes \&  Hernquist  1991) --- which can
make   galaxies  more prone  to starburst  activity  whenever they are
located in denser  environments (e.g., Maia et  al. 1994), at least in
cases of  not extreme environments (Hashimoto  et al.  1998, Balogh et
al. 1998). 

v) Our results supports the fact the normalization of the local galaxy
LF is  relatively low.  Hence,  our local galaxy LF underpredicts many
observed galaxy number  counts (including those  of E/S0 galaxies)  at
relatively bright optical  and near-infrared  magnitudes (where little
galaxy evolution is  in general expected and  observed).  If we do not
want to raise serious doubts on the basic aspects of our understanding
of  galaxy  evolution  and spectral    energy distribution,  the  most
plausible explanation for  this  underprediction  is that the   nearby
volume  we investigate  is  underdense  in galaxies  (by   a factor of
$\sim$1.5).   However, we are  aware  of  the systematic uncertainties
which in general tend to bias low the estimate of the LF normalization
factor (e.g., Willmer 1997) and of the appreciable sensitivity of this
factor to density anomalies within local volumes. 

Notably, uncertainties  related   to the  choice   of  the global   LF
normalization have a much larger impact on the census  of light in the
universe (hence, on  the  determination of the critical  mass-to-light
ratio) than uncertainties in the light contribution of uncataloged LSB
galaxies (see  Dalcanton, Spergel \&   Summers 1997b for  a dissenting
view).

In conclusion,  large observational  efforts (e.g.,  redshift surveys)
are  still needed  to    reliably establish  the  existence,   density
contrast, and  size of a possible  local void,  which, as suggested by
some  authors, can  be larger than   the $\sim$5000 \ks radius  region
examined in this paper.

\begin{center}
{\bf  Acknowledgments}
\end{center}

We wish to thank S.  Bardelli, M. Girardi, G.  Giudice, S. J.  Maddox,
F.   Mardirossian,   M. Mezzetti,    M.    Ramella   for   interesting
conversations and also M. A. Strauss for helpful comments. 
 
In this  work we    have  made use of  the   Lyon-Meudon Extragalactic
Database (LEDA) supplied by the LEDA team  at the CRAL-Observatoire de
Lyon (France). 

P. M. has been supported by  the EC contract ERB
FMB ICT961709. B. C. has been supported by a CIRAC fellowship. C. M. 
thanks SISSA for its kind hospitality. 

This work has been partially   supported by  the Italian Ministry   of
University,  Scientific and Technological  Research (MURST) and by the
Italian Space Agency (ASI).

%%%%%%%%%%%% per formato preprint
%\end{multicols}
%\begin{multicols}{2}
\small
%%%%%%%%%%%% per formato preprint

\end{document}

%% file: tab1.tex
\begin{tabular}{lcccccc}
\hline
\hline
Model    &                                                    
$\alpha$          &
$M_{B}^{*}-5 \log h_{75}$  &
$\phi^{*}$ &
$\chi^{2}/dof$ &
$\rho_{L}$ &
$n$  \\
     &                                                  
    &
    &
$(10^{-3}h_{75}^{3}Mpc^{-3})$ &
  &
$(10^{8}\;L_{\odot}h_{75}\;Mpc^{-3})$&
$(10^{-2}h_{75}^{3}Mpc^{-3})$\\
\hline

Multi--attractor (Mark III)      &$-1.10\pm 0.06$&$-20.67\pm0.08$&$ 5.9 \pm 0.9$& 0.52  &  $1.78\pm0.28$ & $3.2\pm 0.6$ \\

Multi--attractor (Mark III$^{*}$)&$-1.07\pm 0.06$&$-20.70\pm0.08$&$ 5.6 \pm 0.9 $& 0.42  & $1.67\pm0.29$ & $2.9\pm 0.5$  \\

Cluster dipole model             &$-1.15\pm 0.05$&$-20.63\pm0.08$&$ 5.9 \pm 0.9 $& 0.54  & $1.75\pm0.30$& $ 3.7\pm0.8 $\\ 

Modified cluster dipole model    &$-1.09\pm 0.07$&$-20.82\pm0.08$&$ 5.8 \pm 0.9$& 0.68  &  $1.76\pm0.33$ & $3.2\pm0.7 $\\

Hubble Flow (CMB)                &$-1.11\pm 0.06$&$-20.63\pm0.07$&$ 6.2 \pm 1.0$& 0.49  &  $1.80\pm0.29$& $3.5\pm0.7  $\\ 

Hubble Flow (LG)                 &$-1.15\pm 0.05$&$-20.64\pm0.08$&$ 5.9\pm 0.9 $& 0.24  &   $1.78\pm0.31$ & $3.7\pm 0.8$ \\
\hline
\hline
\end{tabular}

%% file: tab2.tex
\begin{tabular}{lcrrrcr}
\hline
\hline                                                                   
Model     &                                                         
Sample    &
Ngal      &
$\alpha$  & 
$M_{B}^{*}-5 \log h_{75} $ &
$\phi^{*}(10^{-3}h_{75}^{3}Mpc^{-3})$ &
$\chi^{2}/dof $  \\

\hline
                                 &E       &344 &$-0.47\pm0.22$&$-20.75\pm0.26$&$0.46\pm0.12$&$0.56$\\
                                 &S0      &596 &$-1.17\pm0.20$&$-20.44\pm0.26$&$0.81\pm0.20$&$0.37$\\                      
				 &E-S0    &940 &$-0.97\pm0.14$&$-20.69\pm0.18$&$1.03\pm0.24$&$0.51 $\\  
Multi-attractor (Mark III)       &Sa-Sb   &1521&$-0.62\pm0.11$&$-20.51\pm0.12$&$2.20\pm0.46$&$0.34 $\\
                                 &Sc-Sd   &2240&$-0.89\pm0.10$&$-20.39\pm0.11$&$3.12\pm0.59$&$0.44 $\\
                                 &Sm-Im  &619 &$-2.41\pm0.28$&$-21.11\pm0.72$&$0.07\pm0.07$&$0.68 $\\              
\vspace{3mm}
                                 &S-Im  &4380&$-1.10\pm0.07$&$-20.63\pm0.09$&$4.58\pm0.73$&$0.63 $\\

                                 &E       &345 &$-0.56\pm0.22$&$-20.71\pm0.26$&$0.45\pm0.12$&$0.39$\\
                                 &S0      &605 &$-1.03\pm0.21$&$-20.20\pm0.24$&$1.03\pm0.26$&$0.31$\\
				 &E-S0    &950 &$-1.03\pm0.14$&$-20.67\pm0.19$&$1.06\pm0.25$&$0.64 $\\  
Cluster Dipole Model             &Sa-Sb   &1563 &$-0.73\pm0.11$&$-20.48\pm0.12$&$2.24\pm0.47$&$0.49 $\\
                                 &Sc-Sd   &2289 &$-0.97\pm0.09$&$-20.35\pm0.12$&$3.17\pm0.60$&$1.09 $\\
                                 &Sm-Im &597 &$-2.45\pm0.32$&$-21.12\pm0.73$&$0.07\pm0.07$&$0.64 $\\
\vspace{3mm}
				 &S-Im  &4449&$-1.17\pm0.08$&$-20.60\pm0.09$&$4.52\pm0.72$&$0.89 $\\

                                 &E       &346 &$-0.55\pm0.22$&$-20.73\pm0.27$&$0.45\pm0.11$&$0.77$\\
                                 &S0      &589 &$-1.08\pm0.21$&$-20.18\pm0.24$&$0.79\pm0.20$&$0.27$\\
				 &E-S0    &935 &$-1.12\pm0.13$&$-20.72\pm0.19$&$0.95\pm 0.23$&$0.74 $\\  
Hubble Flow in LG frame          &Sa-Sb   &1539&$-0.78\pm0.11$&$-20.57\pm0.13$&$2.10\pm0.44$&$0.57 $\\
                                 &Sc-Sd   &2249&$-0.93\pm0.09$&$-20.33\pm0.11$&$3.21\pm0.61$&$0.37 $\\
                                 &Sm-Im   &620&$-2.27\pm0.32$&$-20.75\pm0.69$&$0.15\pm0.15$&$0.42 $\\
				 &S-Im  &4408&$-1.17\pm0.07 $&$-20.60\pm0.09$&$4.62\pm0.70$&$0.45 $\\

\hline
\hline
\end{tabular}

%% file: tab3.tex
\begin{tabular}{lcrrrr}
\hline
\hline

Model    &                                                         
Sample &
Ngal &
$ \alpha $ & 
$M^{*}-5 \log h_{75}$ &
$\chi^{2}/dof$ \\
\hline

				 &field        &2652&$-1.19\pm0.10 $&$-20.59\pm0.12 $&$0.23 $\\  
                                 &groups       &2668&$-1.02\pm0.07 $&$-20.77\pm0.10 $&$0.58 $\\
Multi-attractor (Mark III)       &groups($n\geq10$)  &963 &$-1.21\pm0.11 $&$-20.99\pm0.18 $&$0.58 $\\
\vspace{3mm}
                                 &groups($n\geq20$)  &394 &$-1.28\pm0.18 $&$-21.00\pm0.31$&$0.28$\\

                                 &field        &2735&$-1.24\pm0.09 $&$-20.57\pm0.12 $&$0.59 $\\
                                 &groups       &2664&$-1.07\pm0.08 $&$-20.69\pm0.11 $&$0.38 $\\
Cluster dipole model             &groups($n\geq10$) &937 &$-1.20\pm0.12 $&$-20.95\pm0.23 $&$0.62 $\\
\vspace{3mm}

                                 &groups($n\geq20$) &416 &$-1.07\pm0.17 $&$-20.83\pm0.28 $&$0.49 $\\

                                 &field        &2637&$-1.18\pm0.09 $&$-20.63\pm0.12 $&$0.25 $\\ 
                                 &groups       &2706&$-1.15\pm0.07 $&$-20.70\pm0.11 $&$0.43 $\\
Hubble Flow in LG frame          &groups($n\geq10$)  &952 &$-1.35\pm0.11 $&$-20.97\pm0.19 $&$1.06 $\\
                                 &groups($n\geq20$)  &395 &$-1.37\pm0.17 $&$-21.05\pm0.44 $&$0.49 $\\
\hline
\hline
\end{tabular}

%% file: tab4.tex
\begin{tabular}{llrlcccc}
\hline
\hline

Sample    &
Reference &
Ngal  &
$m_{lim}$ &
$M_{min}- 5 \log h_{75}$  &                                                      
$\alpha$          &
$M_{B}^{*}-5 \log h_{75}$  &
$\phi^{*}(h_{75}^{3}Mpc^{-3})$ \\

\hline
 Stromlo-APM  & Loveday et al., 1992&$\sim 1700$&$b_j=17.15$&-15.9& $-0.97\pm 0.15$&$-20.42\pm 0.13$&$ (0.59\pm0.07) \cdot 10^{-2} $\\   

 CfA2   & Marzke et al., 1994 & $\sim$ 9100 & $mz=15.5$ & -17.9 & $-1.0\pm 0.2$ & $-20.2\pm 0.3$ & $ (1.7\pm 0.4) \cdot 10^{-2} $\\
 
 LCRS    & Lin et al., 1996&$\sim$18700&Gunn r$\sim 17.4$ &-17.4 & $-0.70\pm 0.05$&$-20.11\pm 0.02$&$ (0.8\pm0.04) \cdot 10^{-2} $\\   

 Autofib & Ellis et al., 1996  &$\sim$ 600 &$b_j=24$& -14.9 &$-1.16\pm 0.05$&$-20.22^{+0.15}_{-0.12} $&$ (1.04^{+0.16}_{-0.13}) \cdot 10^{-2} $  \\

 SSRS2  & Marzke \& da Costa, 1997&$\sim$3300&$B(0)=15.5$&-14.9& $-1.16^{+0.08}_{-0.06}$&$-20.37\pm 0.08$&$ (0.46\pm 0.13) \cdot 10^{-2}$ \\

 ESP     & Zucca et al., 1997 &$\sim$3300&$b_j=19.4$& -13.3&$-1.22^{+0.06}_{-0.07} $&$-20.53^{+0.06}_{-0.08}$&$ (0.8\pm0.02) \cdot 10^{-2} $\\

 CS     & Geller et al., 1997  & $\sim$ 1700 & $R=16.13$ & -15.6 & $-1.17\pm0.19$& $-20.35^{+0.17}_{-0.18}$& $ (1.1\pm0.3) \cdot 10^{-2}$ \\ 

 Durham/UKST  & Ratcliffe et al., 1998 &$\sim 2100$ & $b_j \sim 17$& -15.9 & $-1.04\pm 0.08$&$-20.60\pm 0.10$&$ (0.72\pm0.01) \cdot 10^{-2} $\\
\hline
\hline
\end{tabular}